\documentclass{aa}
\usepackage{graphicx}
\usepackage{txfonts}
\usepackage{natbib}
\usepackage{color}
\bibpunct{(}{)}{;}{a}{}{,}    
\newcommand{\beq}{\begin{equation}}
\newcommand{\eeq}{\end{equation}}
\newcommand{\bea}{\begin{eqnarray}}
\newcommand{\eea}{\end{eqnarray}}

\newcommand{\gsim}{\raisebox{-0.6ex}{$\stackrel{{\displaystyle>}}{\sim}$}}

\newcommand{\url}[1]{{\tt #1}}

\def\gapp{\lower 3pt\hbox{${\buildrel > \over \sim}$}\ }
\def\lapp{\lower 3pt\hbox{${\buildrel < \over \sim}$}\ }

\usepackage{color}
\usepackage{ulem}

\newlength{\linwx}
\begin{document}
\title{Highly inclined and eccentric massive planets I: Planet-disc interactions}
\author{
Bertram Bitsch \inst{1},
Aur\'{e}lien Crida \inst{1},
Anne-Sophie Libert \inst{2,3}
\and
Elena Lega \inst{1}
}
\offprints{B. Bitsch,\\ \email{bertram.bitsch@oca.eu}}
\institute{
University of Nice-Sophia Antipolis / CNRS / Observatoire de la C\^{o}te d'Azur, Laboratoire Lagrange UMR 7293, Boulevard de l'Observatoire, BP4229, 06304 NICE cedex 4, FRANCE
\and
NaXys, Department of Mathematics, University of Namur, 8 Rempart de la Vierge, 5000 Namur, Belgium
\and
Observatoire de Lille (LAL-IMCCE), CNRS-UMR8028, 1 Impasse de l'Observatoire, 59000 Lille, France
}
\abstract
{In the Solar System, planets have a small inclination with respect to the equatorial plane of the Sun, but there is evidence that in extrasolar systems the inclination can be very high. This spin-orbit misalignment is unexpected, as planets form in a protoplanetary disc supposedly aligned with the stellar spin. It has been proposed that planet-planet interactions can lead to mutual inclinations during migration in the protoplanetary disc. However, the effect of the gas disc on inclined giant planets is still unknown.
}
{ In this paper we investigate planet-disc interactions for planets above $1M_{\rm Jup}$. We check the influence of three parameters: the inclination $i$, eccentricity $e$, and mass $M_p$ of the planet. This analysis also aims at providing a general expression of the eccentricity and inclination damping exerted on the planet by the disc.
}
{ We perform three-dimensional numerical simulations of protoplanetary discs with embedded high-mass planets on fixed orbits. We use the explicit/implicit hydrodynamical code {{\tt NIRVANA}} in 3D with an isothermal equation of state. 
} 
{We provide damping formulae for $i$ and $e$ as a function of $i$, $e$, and $M_p$ that fit the numerical data. For highly inclined massive planets, the gap opening is reduced, and the damping of $i$ occurs on time-scales of the order of $10^{-4} {\rm deg}/{\rm year} \cdot M_{\rm disc}/(0.01 M_\star)$ with the damping of $e$ on a smaller time-scale. While the inclination of low planetary masses ($<5M_{Jup}$) is always damped, large planetary masses with large $i$ can undergo a Kozai-cycle with the disc. These Kozai-cycles are damped through the disc in time. Eccentricity is generally damped, except for very massive planets ($M_p \sim 5M_{\rm Jup}$) where eccentricity can increase for low inclinations. So the dynamics tends to a final state: planets end up in midplane and can then, over time, increase their eccentricity as a result of interactions with the disc.
}
{ The interactions with the disc lead to damping of $i$ and $e$ after a scattering event of high-mass planets. If $i$ is sufficiently reduced, the eccentricity can be pumped up because of interactions with the disc. If the planet is scattered to high inclination, it can undergo a Kozai-cycle with the disc that makes it hard to predict the exact movement of the planet and its orbital parameters at the dispersal of the disc.
}
\maketitle
\markboth
{Bitsch et al.: Highly inclined massive planets I: planet-disc interactions}
{Bitsch et al.: Highly inclined massive planets I: planet-disc interactions}

\section{Introduction}
\label{sec:introduction}

In the Solar System, the orbits of all the planets are nearly coplanar (within 4 degrees, except for Mercury). The ecliptic (the plane of the Earth's orbit) is also close to the equatorial plane of the Sun: the spin-orbit misalignment is only $\beta_\oplus=7.5^\circ$. The low inclination of the massive planets with respect to the ecliptic is normally taken as an indication that planets form within a flattened protoplanetary disc, itself closely aligned with the stellar equator. The newly discovered Kepler-30 system \citep{2012Natur.487..449S} is even flatter, and confirms this view. However, exo-planets with strong spin-orbit misalignment have been detected (e.g. $\beta>50^\circ$ \citet{2011A&A...533A.113M, 2011epsc.conf..596M, 2011A&A...527L..11H, 2011MNRAS.414.3023S}). Considering that the plane of the past protoplanetary disc should be identical to the present stellar equator\footnote{This is generally accepted, but is actually the subject of debate (see e.g. \citet{2011EPJWC..1103003C, 2012Natur.491..418B}).}, the orbital plane of these planets must have been changed by some mechanism. 

One process generally invoked to explain inclined orbits is scattering by multiple planets in the system after the protoplanetary disc has dissipated (e.g. \citet{2002Icar..156..570M,2008ApJ...686..580C,2008ApJ...686..603J}). These works assume that unstable crowded systems are formed, and undergo planet-planet scattering after a relatively short time when the gas nebula dissipates. However, recent work suggests that unstable systems reach instability while still embedded in the gas disc \citep{Legaetal2013}. A second process is planet-planet interactions during migration in the protoplanetary disc \citep{2003ApJ...597..566T, 2009MNRAS.400.1373L, 2011MNRAS.412.2353L, 2011CeMDA.111..201L}. During the gas-driven migration, the system can enter an inclination-type resonance or the resonant configuration becomes unstable as the resonance excites the eccentricities of the planets and planet-planet scattering sets in. All this affirms the need for a better understanding of the interactions between giant planets and a gaseous protoplanetary disc when the orbit of the former is highly inclined with respect to the midplane of the later. Here, we study this phenomenon, in detail.

\citet{2004ApJ...602..388T} have shown in linear studies that the inclination of a low-mass planet embedded in a disc is exponentially damped by planet-disc interactions for any non-vanishing inclination. Such results are formally valid only for $i \ll H/r$. However, numerical simulations of more highly inclined planets have shown that the exponential damping might be valid up to $i \approx 2 H/r$. If the planet has an even greater inclination, the damping rates deviate from being exponential and it can be fitted by a $di/dt \propto i^{-2}$ function \citep{2007A&A...473..329C, 2011A&A...530A..41B}. However, for high-mass planets, the linear regime is no longer valid. \citet{0004-637X-705-2-1575} considered Jovian-type planets on inclined and eccentric orbits. They find highly inclined and eccentric planets with Jovian masses lose their inclination and eccentricity very quickly (on a time-scale of the order of $10^3$ years) when entering the disc again  (when $i<H/r$). Since a highly inclined planet is only slightly disturbed by the accretion disc (and vice versa), this kind of planet is only able to open a gap in the disc when the inclination drops to $i<10.0^\circ$.

Planet-disc interactions also influence the eccentricity of embedded planets, as has been shown by \cite{1980ApJ...241..425G}. It has been suggested, by performing linear analysis, that the planetary eccentricity can be increased through planet-disc interaction under some conditions \citep{2003ApJ...585.1024G,  2004ApJ...606L..77S, 2008Icar..193..475M}. They estimate that eccentric Lindblad resonances can cause eccentricity growth for gap-forming planets. However, numerical simulations show that eccentricity in the disc is damped for a variety of masses \citep{2007A&A...473..329C, 2009arXiv0904.3336M, 2010A&A.523...A30}.

For very-high-mass planets, on the other hand an eccentric instability in the disc can arise \citep{2006A&A...447..369K}. In turn, this eccentric disc can possibly increase the planetary eccentricity \citep{2001A&A...366..263P,2006ApJ...652.1698D}. However, this process can only explain the eccentricity of very massive ($\approx 5-10 M_{Jup}$) planets. \citet{2013MNRAS.431.1320X} have recently studied the interactions between Jupiter-mass planets and circumstellar discs as well. However, they did not consider planets on eccentric orbits and they were using SPH simulations, while we use a grid-based code.

In this paper, we investigate planet-disc interactions for planets above $1M_{\rm Jup}$, considering different inclination and eccentricity values. Our analysis also aims at deriving a formula for the change of eccentricity and inclination due to planet-disc interactions, in order to study the long-term evolution of systems with massive planets. Indeed, long-term evolution studies of planetary systems cannot be done with hydrodynamical simulations, as the computation time is too long, and N-Body codes that consider the gravitational effects only are used. A correct damping rate of eccentricity and inclination is needed in order to simulate the evolution correctly. This study will be the topic of our paper~II.

We use isothermal three-dimensional (3D) simulations to determine the change of inclination and eccentricity due to planet disc interactions. In Sect.~\ref{sec:methods} we describe the numerical methods used, as well as the procedure to calculate the forces acting on the embedded planets to determine $di/dt$ and $de/dt$. In Sect.~\ref{sec:change} we show $di/dt$ and $de/dt$ as a function of inclination $i$ and eccentricity $e$, and provide fitting formulae. Additionally an observed oscillatory behaviour is discussed in this section. The implications for single-planet systems are shown in Sect.~\ref{sec:application}.

\section{Physical modelling}
\label{sec:methods}

The protoplanetary disc is modelled as a 3D, non-self-gravitating gas whose motion is described by the Navier-Stokes equations. We use the code {\tt Nirvana} \citep{1997ZiegYork,2001ApJ...547..457K}, which uses the {\tt FARGO}-algorithm \citep{2000A&AS..141..165M} and was described in our earlier work on planets on inclined orbits \citep{2011A&A...530A..41B}. We note that the use of the {\tt FARGO}-algorithm may not be straight forward in the case of  highly inclined planets ($i=75.0^\circ$). Our test simulations, however, show that this algorithm can also be used in highly inclined planets, see Appendix.~\ref{ap:numerics}. Here we treat the disc as a viscous medium in the locally isothermal regime. We do not use radiation transport, as we focus here on high-mass planets that open a gap inside a disc, where the effects of heating and cooling of the disc are much less important than for low-mass planets \citep{2009A&A...506..971K}. A more detailed description of the used code can be found in \citet{2009A&A...506..971K}.

\subsection{Smoothing of the planetary potential}
\label{subsec:pot}

An important issue in modelling planetary dynamics in discs is the gravitational potential of the planet since this has to be artificially smoothed to avoid singularities.  While in two dimensions a potential smoothing takes care of the otherwise neglected vertical extension of the disc, in three dimensional simulations the most accurate potential should be used. As the planetary radius is much smaller than a typical grid cell, and the planet is treated as a point mass, a smoothing of the potential is required to ensure numerical stability.

In \citet{2009A&A...506..971K} two different kinds of planetary potentials for 3D discs have been discussed. The first is the classic $\epsilon_{sm}$-potential 
\begin{equation}
\label{eq:epsilon}
   \Phi_p^{\epsilon_{sm}} = - \frac{G M_p}{\sqrt{d^2 + \epsilon_{sm}^2}} \, .
\end{equation}
Here $M_P$ is the planetary mass, and $d=| \mathbf{r} - \mathbf{r_P}|$ denotes the distance of the disc element to the planet. This potential has the advantage that it leads to very stable evolutions when the parameter $\epsilon_{sm}$ is a significant fraction of the Roche radius. The disadvantage is that for smaller $\epsilon_{sm}$, which would yield a higher accuracy at larger $d$, the potential becomes very deep at the planetary position. Additionally, the potential differs from the exact $1/r$ potential even for medium to larger distances $d$ from the planet. 

To resolve these problems at small and large $d$ simultaneously, the following {\it cubic}-potential has been suggested \citep{2006A&A...445..747K,2009A&A...506..971K} 
\begin{equation}
\label{eq:cubic}
\Phi_p^{cub} =  \left\{
    \begin{array}{cc} 
   - \frac{G M_p}{d} \,  \left[ \left(\frac{d}{r_\mathrm{sm}}\right)^4
     - 2 \left(\frac{d}{r_\mathrm{sm}}\right)^3 
     + 2 \frac{d}{r_\mathrm{sm}}  \right]
     \quad &  \mbox{for} \quad  d \leq r_\mathrm{sm} \ \textcolor{white}{.} \\
   -  \frac{G M_p}{d}  \quad & \mbox{for} \quad  d > r_\mathrm{sm} \ .
    \end{array}
    \right.
\end{equation}
The construction of the planetary potential is such that for distances larger than $r_{sm}$ the potential matches the correct $1/r$ potential. Inside this radius ($d < r_{sm}$) it is smoothed by a cubic polynomial. This potential has the advantage of exactness outside the specified distance $r_{sm}$, while being finite inside. 

For $1M_{\rm Jup}$ and $5M_{Jup}$ we use the cubic potential with $r_{sm}=0.8 R_H$. For the $10M_{\rm Jup}$ planet, we use the $\epsilon_{sm}$-potential with $r_{sm}=0.8 R_H$, with the Hill radius $R_H$ given by
\begin{equation}
 R_{H} = a_p \left(\frac{M_p}{3 M_\star}\right)^{1/3} \ ,
\end{equation}
where $a_p$ is the semi major axis of the planet, and $M_\star$ is the mass of the central star.

As the planetary mass increases, so does the amount of material accumulated near the planet. In order to resolve the gradients of density in that region correctly, a much higher resolution is required. Therefore, we change the cubic potential to the $\epsilon_{sm}$-potential for the $10M_{\rm Jup}$ planet. For the torque acting on the planets, the consequences are minimal, as we use a torque cut-off function in the Hill sphere of the planet, as described below. Additional information regarding the smoothing length can be found in Appendix~\ref{ap:numerics}.

\subsection{Initial setup}

The three-dimensional ($r, \theta, \phi$) computational domain consists of a complete annulus of the protoplanetary disc centred on the star, extending from $r_{min}=0.2$ to $r_{max}=4.2$ in units of $r_0=a_{Jup}=5.2 AU$, where we put the planet. The planet is held on a fixed orbit during the evolution. The eccentricity of the planet can be $e_0=0.0$, $e_0=0.2$, or $e_0=0.4$. We use $390 \times 48 \times 576$ active cells for the simulations with $1M_{Jup}$ and $260 \times 32 \times 384$ active cells for $5M_{Jup}$ and $10M_{Jup}$. This resolution is sufficient, as we still resolve the horseshoe width with a few grid cells for all planetary masses. The horseshoe width is defined for large planets as $x_s = \sqrt{12} a_P (q/3)^{1/3}$ \citep{2006ApJ...652..730M}, where $q$ is the planet-star mass ratio. Tests regarding the numerical resolution can be found in Appendix~\ref{ap:numerics}.

In the vertical direction, the annulus extends $7^\circ$ below and above the disc's midplane, meaning $83^\circ < \theta < 97^\circ$. Here $\theta$ denotes the polar angle of our spherical polar coordinate system measured from the polar axis, therefore the midplane of the disc is at $\theta=90.0^\circ$. We use closed boundary conditions in the radial and vertical directions. In the azimuthal direction, periodic boundary conditions are used. The central star has one solar mass $M_\ast = M_\odot$, and the total disc mass inside [$r_{min}, r_{max}$] is $M_{disc} = 0.01 M_\odot$. The aspect ratio of the disc is $H/r=0.05$. We use an $\alpha$ prescription of the viscosity, where $\nu = \alpha c_s^2 / \Omega_K$ \citep{1973A&A....24..337S} with $\alpha=0.005$\,; $\Omega_K$ is the Kepler frequency\,; $c_s = \sqrt{P/\rho}$ denotes the isothermal sound speed, $P$ the pressure, $\rho$ the volume density of the gas, and $H=c_s/\Omega$. 

The models are initialised with constant temperatures on cylinders with a profile $T(s) \propto s^{-1}$ with $s=r \sin \theta$. This yields a constant ratio of the disc's vertical height $H$ to the radius $s$. The initial vertical density stratification is given approximately by a Gaussian
\begin{equation}
  \rho(r,\theta)= \rho_0 (r) \, \exp \left[ - \frac{(\pi/2 - \theta)^2 \, r^2}{2 H^2} \right] \ .
\end{equation} 
Here, the density in the midplane is $\rho_0 (r) \propto r^{-1.5}$ which leads to a $\Sigma(r) \propto \, r^{-1/2}$ profile of the vertically integrated surface density. In the radial and $\theta$-direction we set the initial velocities to zero, while for the azimuthal component the initial velocity $u_\phi$ is given by the equilibrium of gravity, centrifugal acceleration and the radial pressure gradient. This corresponds to the equilibrium configuration for a purely isothermal disc with constant viscosity. However, as the massive planets in the disc start to open gaps, the density and surface density profile get distorted.

\subsection{Calculation of forces}
\label{subsec:forces}

To determine the change of orbital elements for planets on fixed inclined orbits, we follow \citet{burns:944} and compute the forces as described in \citet{2011A&A...530A..41B}. The gravitational torques and forces acting on the planet are calculated by integrating over the whole disc, where we apply a tapering function to exclude the inner parts of the Hill sphere of the planet. Specifically, we use the smooth (Fermi-type) function
\begin{equation}
\label{eq:fermi}
     f_b (d)=\left[\exp\left(-\frac{d/R_H-b}{b/10}\right)+1\right]^{-1}
\end{equation}
which increases from 0 at the planet location ($d=0$) to 1 outside $d \geq R_{H}$ with a midpoint $f_b = 1/2$ at $d = b R_{H}$, i.e. the quantity $b$ denotes the torque-cutoff radius in units of the Hill radius. This torque-cutoff is necessary to avoid large, probably noisy contributions from the inner parts of the Roche lobe and to disregard material that is possibly gravitationally bound to the planet \citep{2009A&A...502..679C}. Here we assume $b= 0.8$, as a change in $b$ did not influence the results significantly \citep{2009A&A...506..971K}.

If a small disturbing force $\mathbf{dF}$ (given per unit mass) due to the disc is acting on the planet, the planet changes its orbit. This small disturbing force $\mathbf{dF}$ may change the planetary orbit in size (semi-major axis $a$), eccentricity $e$, and inclination $i$. The inclination $i$ gives the angle between the orbital plane and an arbitrary fixed plane, which is in our case the equatorial plane ($\theta = 90^\circ$), which corresponds to the midplane of the disc. Only forces lying in the orbit plane can change the orbit's size and shape, while these forces cannot change the orientation of the orbital plane. In \citet{burns:944} the specific disturbing force is written as
\begin{equation}
\label{eq:distforce}
     \mathbf{dF} = \mathbf{R} + \mathbf{T} + \mathbf{N} = R \mathbf{e}_{R} + T \mathbf{e}_T + N \mathbf{e}_N \ ,
\end{equation}
where each $\mathbf{e}$ represents the relevant orthogonal component of the unit vector. The perturbing force can be split into these components: $\mathbf{R}$ is radially outwards along $\mathbf{r}$; $\mathbf{T}$ is transverse to the radial vector in the orbit plane (positive in the direction of motion of the planet); and $\mathbf{N}$ is normal to the orbit planet in the direction $\mathbf{R} \times \mathbf{T}$.

\citet{burns:944} finds for the change of inclination
\begin{equation}
\label{eq:ichange}
     \frac{di}{dt} = \frac{a N \cos \xi}{H} \ ,
\end{equation}
where the numerator is the component of the torque which rotates the specific angular momentum $\mathbf{H} = \mathbf{r} \times \mathbf{\dot{r}}$ about the line of nodes (and which thereby changes the inclination of the orbital plane). The specific angular momentum $H$ is defined as
\begin{equation}
	H = \sqrt{ G M_\star a_p (1 - e^2)} \ .
\end{equation}
The angle $\xi$ is related to the true anomaly $f$ by $f = \xi - \omega$, with $\omega$ being the argument of periapsis and $\xi$ describes the angle between the line of nodes and the planet on its orbit around the star. For the case of circular orbits, the argument of periapsis $\omega$ is zero.

The change of eccentricity is given by \citet{burns:944} as
\begin{equation}
\label{eq:eccchange}
  \frac{de}{dt} = \left[ \frac{a (1-e^2)}{GM_\star} \right]^{1/2} \left[R \sin f + T (\cos f + \cos \epsilon) \right] \ ,
\end{equation}
where $\epsilon$ is the eccentric anomaly, which is given by
\begin{equation}
 \cos \epsilon = \frac{e + \cos f}{1 + e \cos f} \ .
\end{equation}
With this set of equations, we are able to calculate the forces acting on planets on fixed orbits and determine $di/dt$ and $de/dt$.

\section{Planets on inclined and eccentric orbits}
\label{sec:change}

In this section we investigate the changes of the planetary orbit due to planet-disc interactions. The planets are put in fixed orbits with inclinations ranging from $i_0=1.0^\circ$ to $i_0=75^\circ$, with a total of ten different inclinations. For each inclination we also adopt three different eccentricities, which are $e_0=0.0$, $e_0=0.2$ and $e_0=0.4$. 

We note that the orbit of highly inclined planets is not embedded completely in the hydrodynamical grid, since the grid is only extended up to $7^\circ$ above and below midplane. However, the density distribution in the vertical direction follows a Gaussian profile and for an aspect ratio of $0.05$ we are at about $2.5\sigma$ at $7^\circ$ so that the contribution of the gas can be neglected at larger $\theta$.

\subsection{Gaps in discs}

\begin{figure}
 \centering
 \resizebox{\hsize}{!}{\includegraphics[width=0.9\linwx]{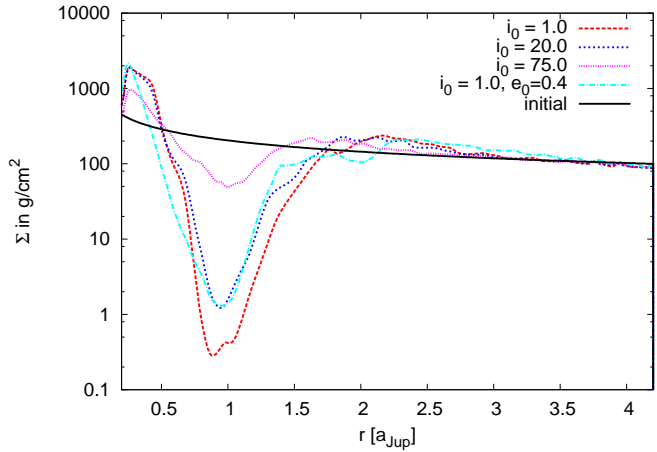}}
 \caption{Surface density for disc simulations with $10M_{Jup}$ planets in circular and eccentric orbits with different inclinations. The surface density is plotted after $400$ planetary orbits. The evolution has reached an equilibrium state, meaning that the surface density does not change in time any more.
   \label{fig:Sig10MJup}
   }
\end{figure}

The criterion for gap opening depends on the viscosity, the pressure, and the planetary mass \citep{2006Icar..181..587C}. Giant planets ($M \gsim 0.5 M_{\rm Jup}$) are generally massive enough to split the disc. However, the inclination of a giant planet plays a very important role in opening a gap as well, as can be seen in Fig.~\ref{fig:Sig10MJup}, where we display the surface density profile of discs with embedded $10M_{\rm Jup}$ planets on different inclinations.

Clearly, a lower inclination produces a much wider and deeper gap inside the disc. For larger inclinations, the gap opening is reduced, as the planet spends less and less time inside the disc to push material away from its orbit. Additionally, eccentric planets open up gaps that are less deep than their circular counter parts. This effect is very important for the damping of inclination and eccentricity, as an open gap inside the disc prolongs the damping time-scale of inclination \citep{2011A&A...530A..41B} and of eccentricity \citep{2010A&A.523...A30}. Gap opening also indicates that linear analysis of the situation is no longer applicable.

\begin{figure}
 \centering
 \resizebox{\hsize}{!}{\includegraphics[width=1.0\linwx]{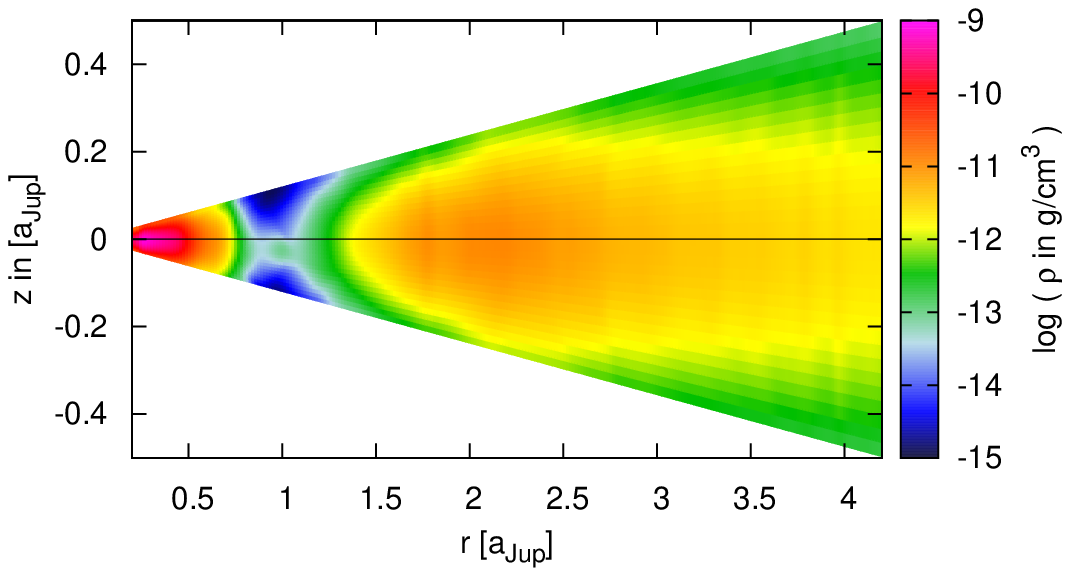}}
 \resizebox{\hsize}{!}{\includegraphics[width=1.0\linwx]{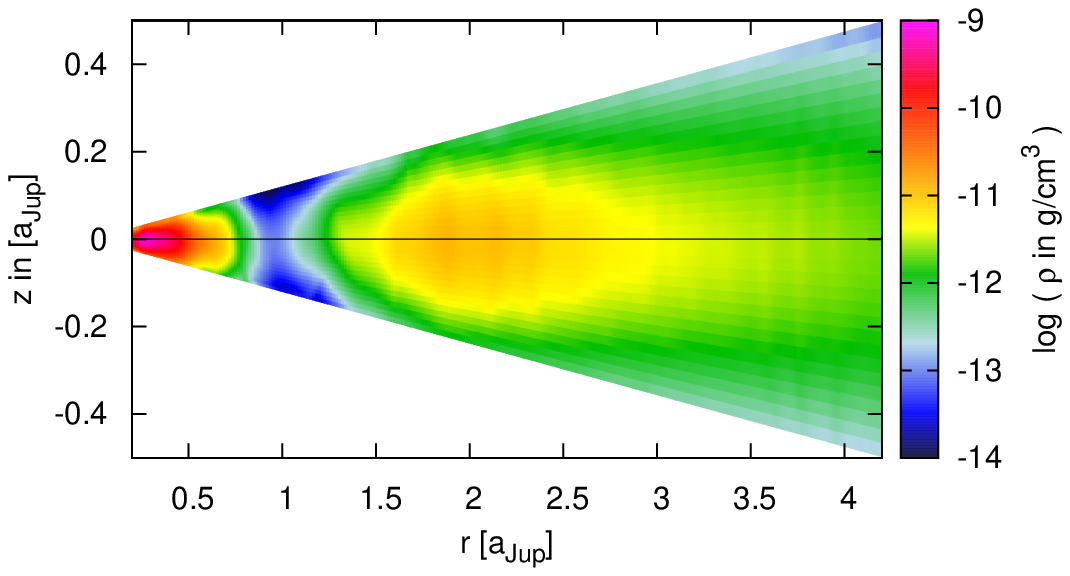}}
 \resizebox{\hsize}{!}{\includegraphics[width=1.0\linwx]{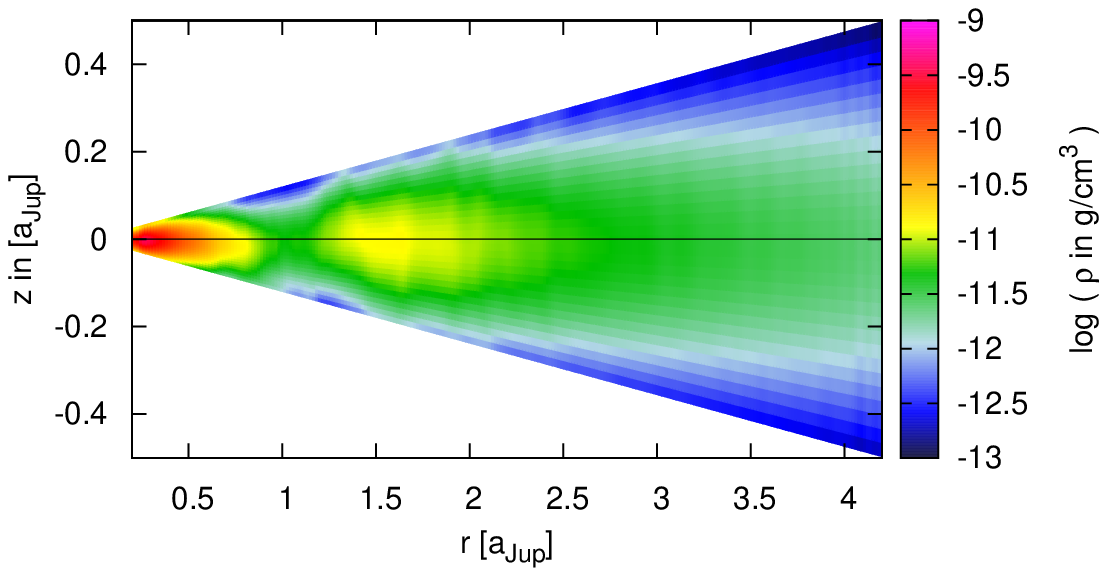}}
 \caption{Density (in $g/cm^3$) of a $r-\theta$-slice through the disc at the azimuth of an embedded $10 M_{Jup}$ planet on a fixed circular inclined orbit with $i_0=1.0^\circ$ (top), $i_0=20^\circ$ (middle), and $i_0=75.0^\circ$ (bottom). The planet is at its lowest point in orbit (lower culmination) at the time of the snapshot, which was taken after $400$ planetary orbits. We note the slightly different colour scale for each plot. The black line indicates the midplane of the grid to which the inclination of the disc is measured (see Sect.~\ref{subsec:eichange}).
   \label{fig:10MJupxyrho}
   }
\end{figure}

In Fig.~\ref{fig:10MJupxyrho} we present slices in the $x-z$-plane for the disc's density for $10M_{Jup}$ planets on inclinations of $1^\circ$, $20^\circ$, and $75^\circ$ degrees. The inclinations correspond to those shown in the surface density plot (Fig.~\ref{fig:Sig10MJup}). Clearly the depth of the gap shown in the surface density is reflected in the 2D plots. Additionally, the density structures show no effects at the upper and lower boundaries because of boundary conditions, indicating that an opening angle of $7^\circ$ is sufficient for our simulations.

\subsection{Change of the disc structure}
\label{subsec:eichange}

It has been known since several years that massive planets are able not only to open up a gap in the disc, but are also able to change the shape of the whole disc by turning it eccentric \citep{2001A&A...366..263P, 2006A&A...447..369K}. Additionally, the inclination of the disc will change due to the interactions with the inclined planet. In this section, we discuss the impact of a massive planet on the eccentricity and inclination of the disc.

\begin{figure}
 \centering
 \resizebox{\hsize}{!}{\includegraphics[width=1.0\linwx]{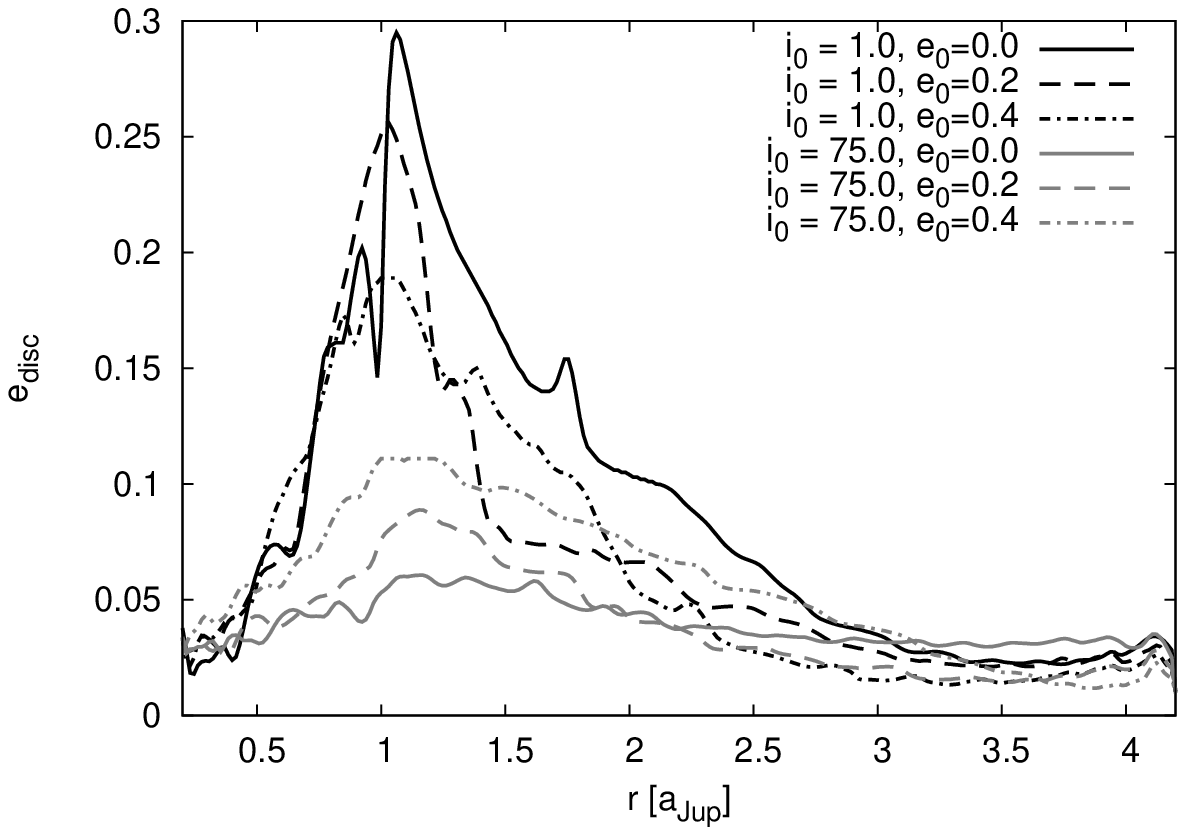}}
 \resizebox{\hsize}{!}{\includegraphics[width=1.0\linwx]{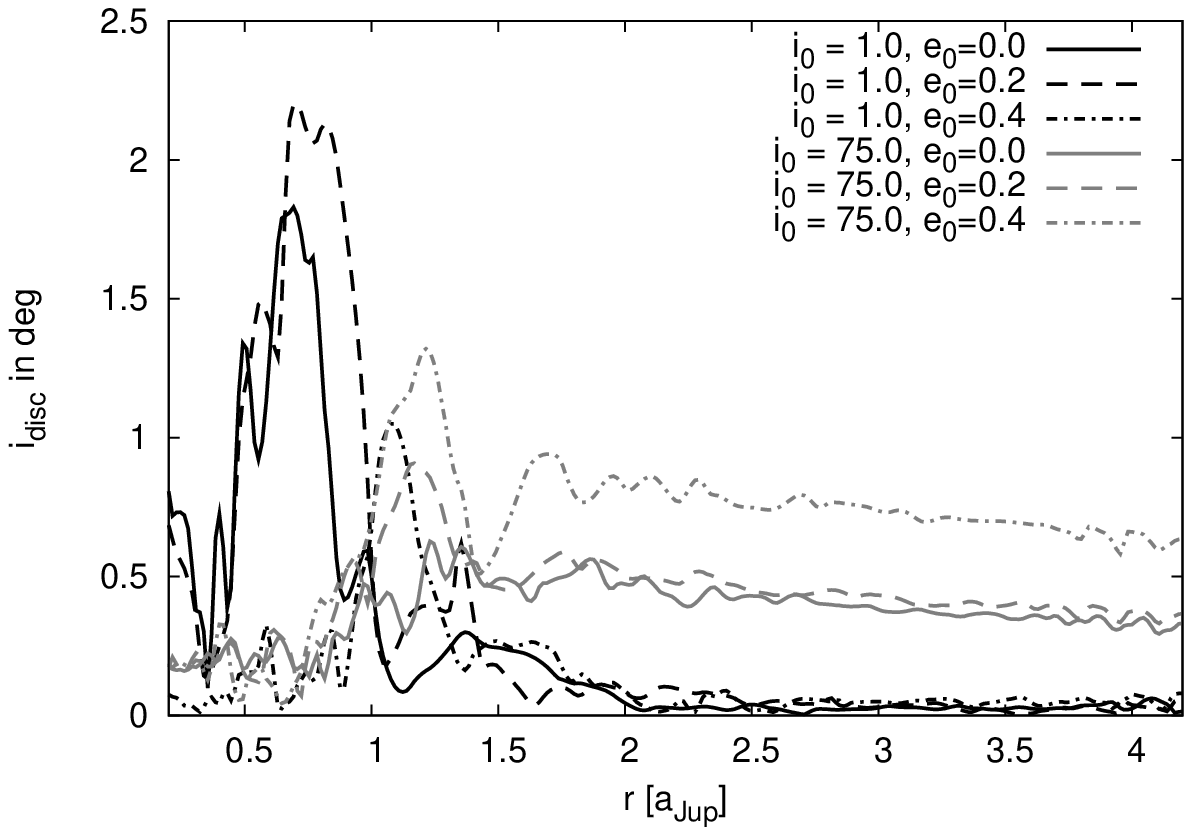}}
 \caption{Eccentricity (top) and inclination (bottom) of the disc with a $10M_{Jup}$ planet influencing the disc structure after $400$ planetary orbits.
   \label{fig:eidisc10MJup}
   }
\end{figure}

In Fig.~\ref{fig:eidisc10MJup} we display the eccentricity (top) and inclination (bottom) of the disc interacting with a $10M_{Jup}$ planet with different inclinations ($1^\circ$ and $75^\circ$) and eccentricities. The calculations for deriving the eccentricity and inclination of the disc can be found in Appendix~\ref{ap:eidisc}.

For low planetary inclinations, the influence of the planet on the eccentricity of the protoplanetary disc is greater than on high planetary inclinations, simply, because the planet is closer to midplane and can therefore influence the eccentricity of the disc more strongly by pushing the material away. The eccentricity increase of the disc is stronger for planets in circular orbits than for planets that are already in an eccentric orbit. For highly inclined planets, the situation is reversed. The disc is most eccentric for planets that are already in an eccentric orbit and the disc is less eccentric for planets in circular orbits. Additionally, the eccentricity of the disc is highest close to the planet and drops with distance from the planet, independent of the inclination of the planet.

The inclination of the disc for the $i_0=1^\circ$ planets is greater mostly around the planet's location (at $r=1.0a_{Jup}$) because the influence of the planet is strongest there. The inclination of the disc can be larger than the inclination of the planet. This is possible because the planet opens a gap inside the disc and pushes the material away from the planet (Fig.~\ref{fig:Sig10MJup}, top), which can also be seen in the 2D density configuration (top of Fig.~\ref{fig:10MJupxyrho}). In the outer parts of the disc the disc remains non-inclined.

For planets with high inclinations the situation is slightly different than for planets with low inclinations. The maximum of inclination is lower and there is no distinct maximum of inclination visible inside the planetary orbit ($r<1.0a_{Jup}$) compared to the case of low inclinations. However, the outer parts of the disc show a non-zero inclination (which has a tendency to be larger for larger planetary eccentricities), which was not visible for the low-inclination planets. Additionally, they show a small peak of inclination at $r \approx 1.25a_{Jup}$.

\subsection{Change of orbital parameters}

\subsubsection{Eccentricity}
\label{subsubsec:Eccentricity}

As stated in Sect.~\ref{subsec:forces}, the forces acting on a planet on a fixed orbit can be calculated and then used to determine a rate of change for the inclination and eccentricity. The damping rates are taken when the planet-disc interactions are in an equilibrium state and do not change on average any more. The damping given by Eq.~(\ref{eq:ichange}) varies strongly within the time of an orbit and slightly from one orbit to an other. Thus, we averaged it over $40$ planetary orbits.

\begin{figure}
 \centering
 \resizebox{\hsize}{!}{\includegraphics[width=0.9\linwx]{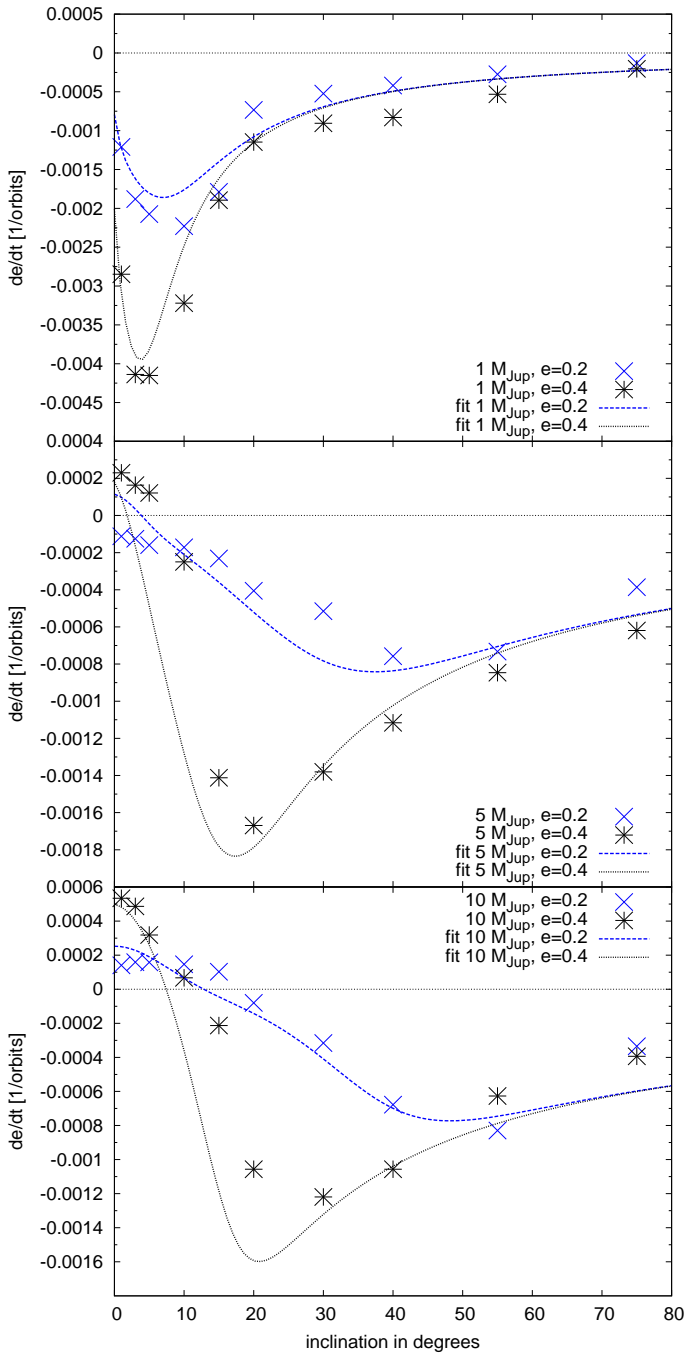}}
 \caption{Change of eccentricity $de/dt$ for planets with $1M_{Jup}$, $5M_{Jup}$, and $10M_{Jup}$ with different eccentricities. Points are results from numerical simulations, while lines indicate the fitting of the data. The $1M_{Jup}$ planets have been evolved for $200$ planetary orbits, the $5M_{Jup}$ and $10M_{Jup}$ planets have been evolved for 400 orbits. The forces used to calculate the data points have been averaged over $40$ planetary orbits for all simulations.
   \label{fig:eccentricity}
   }
\end{figure}

In Fig.~\ref{fig:eccentricity} we present the change of eccentricity $de/dt$ for planets of $1M_{Jup}$, $5M_{Jup}$, and $10M_{Jup}$ on orbits with different inclinations and eccentricities. The change of $de/dt$ has been studied in the past for coplanar low-mass planets \citep{2007A&A...473..329C, 2010A&A.523...A30} and for high-mass planets \citep{2001A&A...366..263P, 2006A&A...447..369K}.

For low inclinations ($i_0<10^\circ$) the damping of eccentricity is stronger than for larger inclinations in the case of $1M_{Jup}$. The maximum damping rate is also dependent on the initial eccentricity $e_0$, where a larger $e_0$ provides a faster damping. The damping of eccentricity is reduced significantly for larger inclinations $i_0>20^\circ$. As soon as the planet is no longer embedded in the disc, the damping reduces, as it is most efficient when the planet is inside the disc and not high above or below the disc for most of its orbit.

For low inclinations ($i_0<10^\circ$) and low eccentricities ($e_0<0.2$), the $5M_{Jup}$ planet opens up a large gap inside the disc. As the planet opens a gap inside the disc the damping is reduced because there is less material close to the planet to damp its orbit. For large initial eccentricities ($e_0=0.4$), an increase of eccentricity is observable for low planetary inclinations. But for higher inclinations, the damping of eccentricity increases as well, because the planet does not open up such a deep gap (Fig.~\ref{fig:Sig10MJup}). However, for $i_0>40^\circ$ the damping of eccentricity becomes smaller again, because the planet spends less and less time in the midplane of the disc where most of the disc material is, which is responsible for damping. 

For even higher masses ($10M_{Jup}$), we observe an eccentricity increase for low planetary inclinations for all non-zero eccentricities. But for larger inclinations ($i_0>15^\circ$), the eccentricity is damped again. The largest value of damping is at $i_0\approx 30^\circ-50^\circ$, depending on the planet's eccentricity and is then reduced for higher inclinations again, following the trend described for the $5M_{Jup}$ planet. 

For large planets with low inclinations, the eccentricity of the planet can rise, which has been observed in \citet{2001A&A...366..263P} and \citet{2006A&A...447..369K}. \citet{2001A&A...366..263P} stated that if the planet opens up a large gap, the $m=2$ spiral wave at the $1:3$ outer eccentric Lindblad resonance becomes dominant (because the order $1$ resonances lie inside the gap) and induces eccentricity growth. However, they found an eccentricity increase only for $M_P>20M_{Jup}$, while our simulations indicate it clearly already for $M_P>5M_{Jup}$ (Fig.~\ref{fig:eccentricity}, bottom). The differences between their 2D simulations and our 3D simulations can be the cause of the change in the required planetary mass for eccentricity growth.

Additionally, by embedding a high-mass planet inside a disc, the disc can become eccentric, as shown in Sect.~\ref{subsec:eichange}. The disc's eccentricity is dependent on the planet's inclination and slightly dependent on its eccentricity as well (see Fig.~\ref{fig:eidisc10MJup}).

It seems that the coupling between a large disc eccentricity at $r \approx 1-1.5 a_P$ and a large planetary eccentricity (the $i_0=1^\circ$ with $e=0.4$ case) results in a large force on the planet. This effect is increased as the planet in an eccentric orbit opens a small gap leaving more material at that location. This leads then to a greater increase of eccentricity for highly eccentric planets, compared to those with small eccentricity.

\subsubsection{Inclination}
\label{subsubsec:Inclination}

\begin{figure}
 \centering
 \resizebox{\hsize}{!}{\includegraphics[width=0.9\linwx]{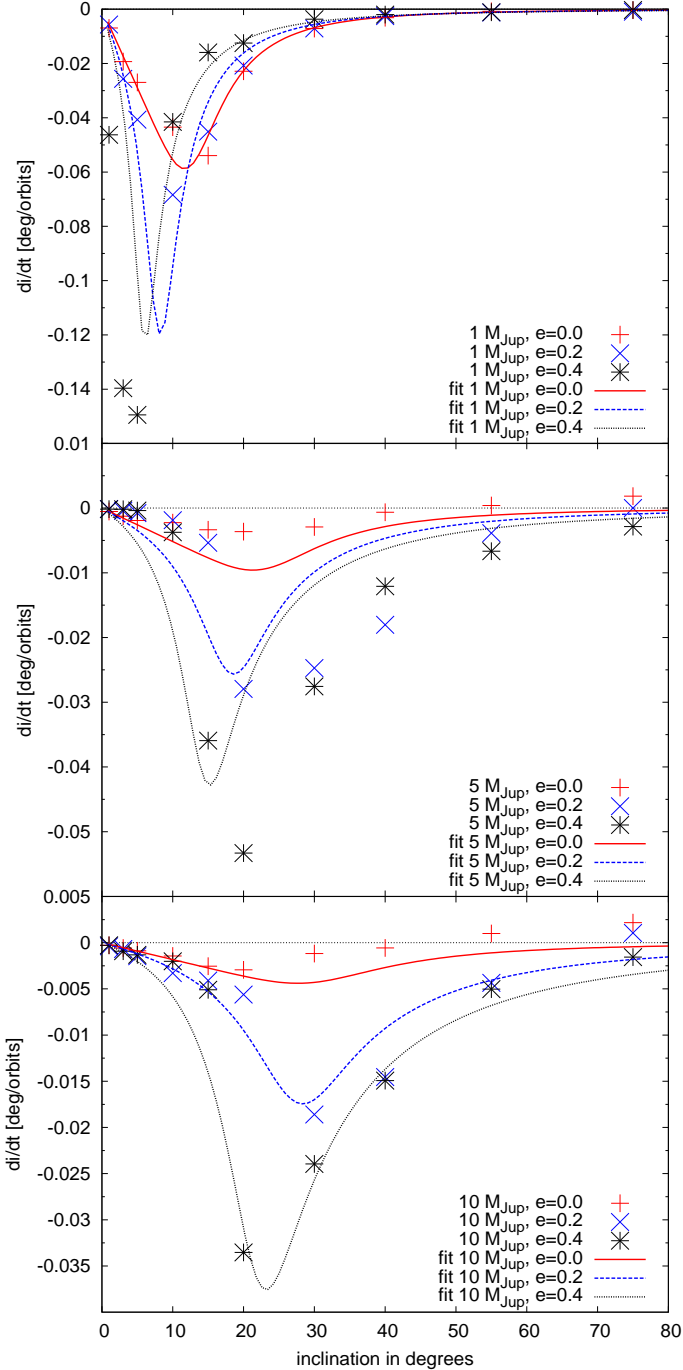}}
 \caption{Change of inclination $di/dt$ for planets with $1M_{Jup}$, $5M_{Jup}$, and $10M_{Jup}$ with different eccentricities. $di/dt$ is in degrees per orbit at the planet's location $r_P=1.0a_{Jup}$. Points are results from numerical simulations, while lines indicate the fitting of the data. The $1M_{Jup}$ planets have been evolved for $200$ planetary orbits, the $5M_{Jup}$ and $10M_{Jup}$ planets have been evolved for $400$ orbits. The forces used to calculate the data points have been averaged over $40$ planetary orbits for all simulations.
   \label{fig:inclination}
   }
\end{figure}

In Fig.~\ref{fig:inclination} we present the rate of change of inclination $di/dt$, presented in degrees per orbit, for planets with different masses and different eccentricities. For $1M_{Jup}$ the inclination is damped for all initial inclinations. For increasing inclinations with $i_0<15^\circ$ (smaller for increasing eccentricity), the damping of inclination increases. This increase is nearly linear, as has been shown for low-mass planets in theory \citep{2004ApJ...602..388T} and in numerical simulations \citep{2007A&A...473..329C, 2011A&A...530A..41B}. The rates of inclination damping for zero-eccentricity planets are comparable to those stated in \citet{2013MNRAS.431.1320X}.

For $i_0>15^\circ$, the damping rate of the inclination is a decreasing function of inclination\,; this is consistent with the planet-disc interaction being weaker when the planet spends more time farther from the midplane.

For $5M_{Jup}$ the damping of inclination is almost the same as for the $1M_{Jup}$ planet, but with a maximum at $i_0 \approx 20^\circ$. However, there is a significant difference for high inclined and low eccentric planets\,: the inclination is not damped if $i_0>50^\circ$, but it increases for $e_0<0.1$. This behaviour will be discussed in Sect.~\ref{subsec:moving}.

The $10M_{Jup}$ planet shows the same general behaviour as the $5M_{Jup}$ planet, but the inclination increase already sets in at $i_0\geqslant 45^\circ$, depending on $e_0$. Still, no inclination increase is observed in the high eccentricity simulations ($e_0=0.4$). We also want to stress here that the damping rate significantly increases with increasing planetary eccentricity for all planetary masses.

The increase of inclination for high-mass planets due to interactions with the disc has been studied in \citet{2001ApJ...560..997L}. They state that the $1:3$ mean-motion resonance also acts to increase inclination. This resonance is at $r_r=2.08r_P$, which clearly is not inside an open gap in the case of $i_0=75^\circ$ (see Fig.~\ref{fig:Sig10MJup}). However, the resonances closer to the planet ($1:2$ and $2:3$) are also not completely inside the gap, so that there should be some damping effects, but the damping of inclination through these resonances is weaker than the increase from the $1:3$ resonance because in total the inclination increases for high inclined planets (Fig.~\ref{fig:inclination}). \citet{2001ApJ...560..997L} also used small $i_0$ for their calculations in order to apply linear theory, which does not apply for large inclinations. The situation for our high inclination planets might therefore be completely different from their calculations.

\subsection{Moving planets in discs}
\label{subsec:moving}

\subsubsection{Short-term evolution}

\begin{figure}
 \centering
 \resizebox{\hsize}{!}{\includegraphics[width=0.9\linwx]{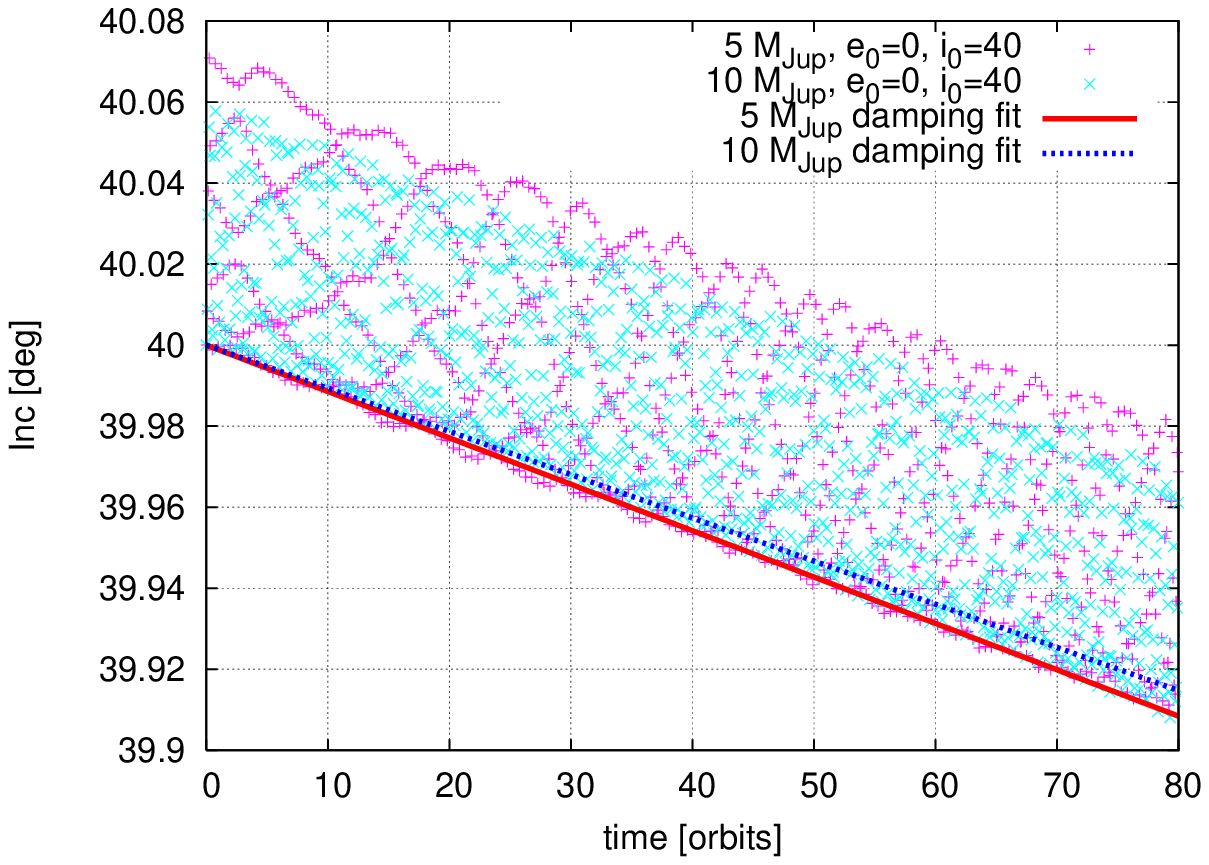}}
 \resizebox{\hsize}{!}{\includegraphics[width=0.9\linwx]{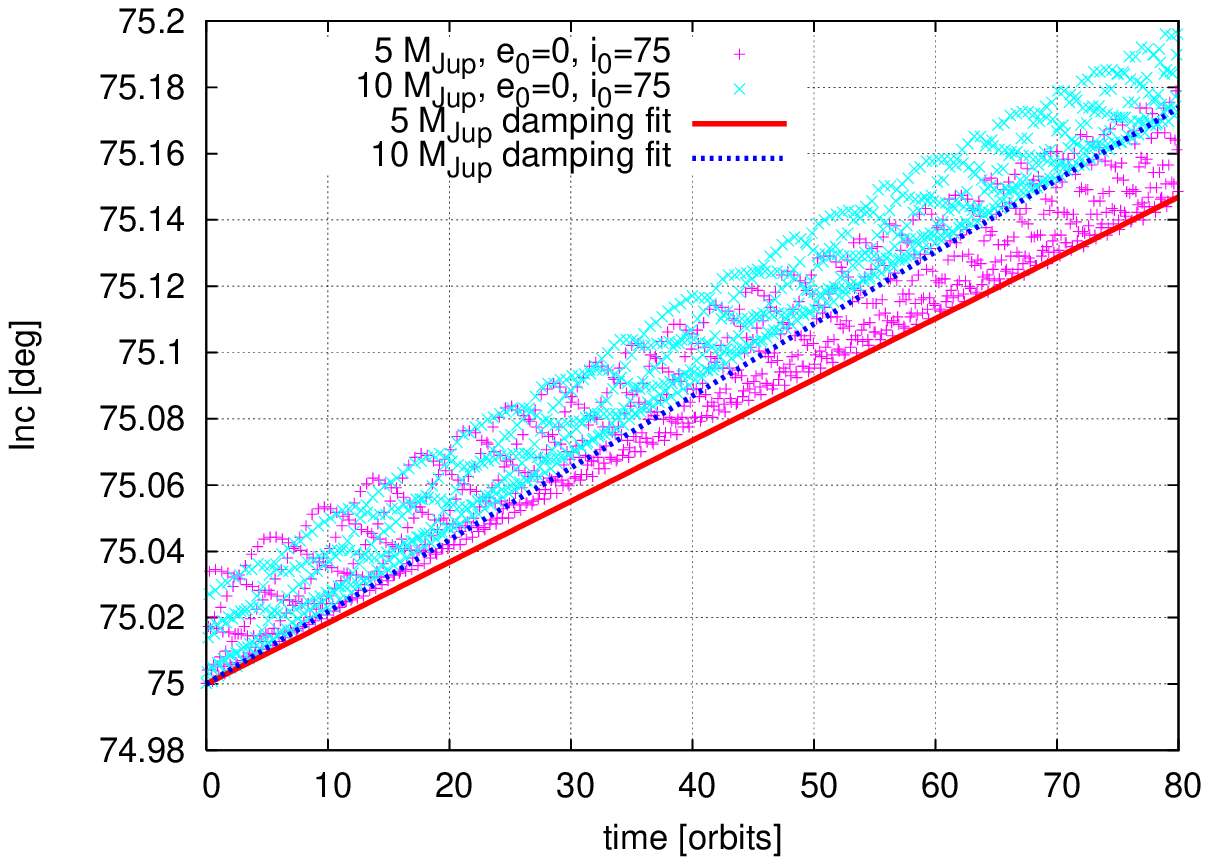}}
 \caption{Evolution of inclination of $5M_{Jup}$ and $10M_{Jup}$ planets with an initial inclination of $i_0=40.0^\circ$ (top) and $i_0=75.0^\circ$ (bottom) in circular orbits. The simulations have been restarted with a moving planet after the disc was evolved for a fixed planet for $400$ planetary orbits. The time index has been reset to zero and the lines correspond to the expected damping rates from Fig.~\ref{fig:inclination}.
   \label{fig:Inc40ev}
   }
\end{figure}

In order to verify the results of inclination and eccentricity change, we present in this section simulations of planets evolving freely in the disc. The planets are moving because of the influences of the discs forces. We present here several interesting cases for planets with high inclinations. The first case is for $5M_{Jup}$ and $10M_{Jup}$ with an inclination of $i_0=40^\circ$ and $i_0=75^\circ$ in circular orbits with an evolution time of $80$ planetary orbits. In Fig.~\ref{fig:Inc40ev} the evolution of inclination with time is presented for the two different planets and inclinations. The evolution is nearly identical, as was predicted by the measured forces for the planets on fixed orbits, which is shown by the solid lines (rates from Fig.~\ref{fig:inclination}).

One should be aware, however, that by keeping the planet in a fixed orbit, angular momentum in the system is not conserved because, for example, the inclination of the disc is rising (see Fig.~\ref{fig:eidisc10MJup}) while the planet remains in a fixed orbit. The effect of conserving angular momentum is not a problem for low-mass planets, where the measured forces match perfectly with the inclination damping rates for moving planets \citep{2011A&A...530A..41B}, but for big planets of several Jupiter masses this can lead to small differences because the angular momentum transfer from disc to planet and vice-versa is much larger.

\subsubsection{Long-term evolution}

\begin{figure}
 \centering
 \resizebox{\hsize}{!}{\includegraphics[width=0.9\linwx]{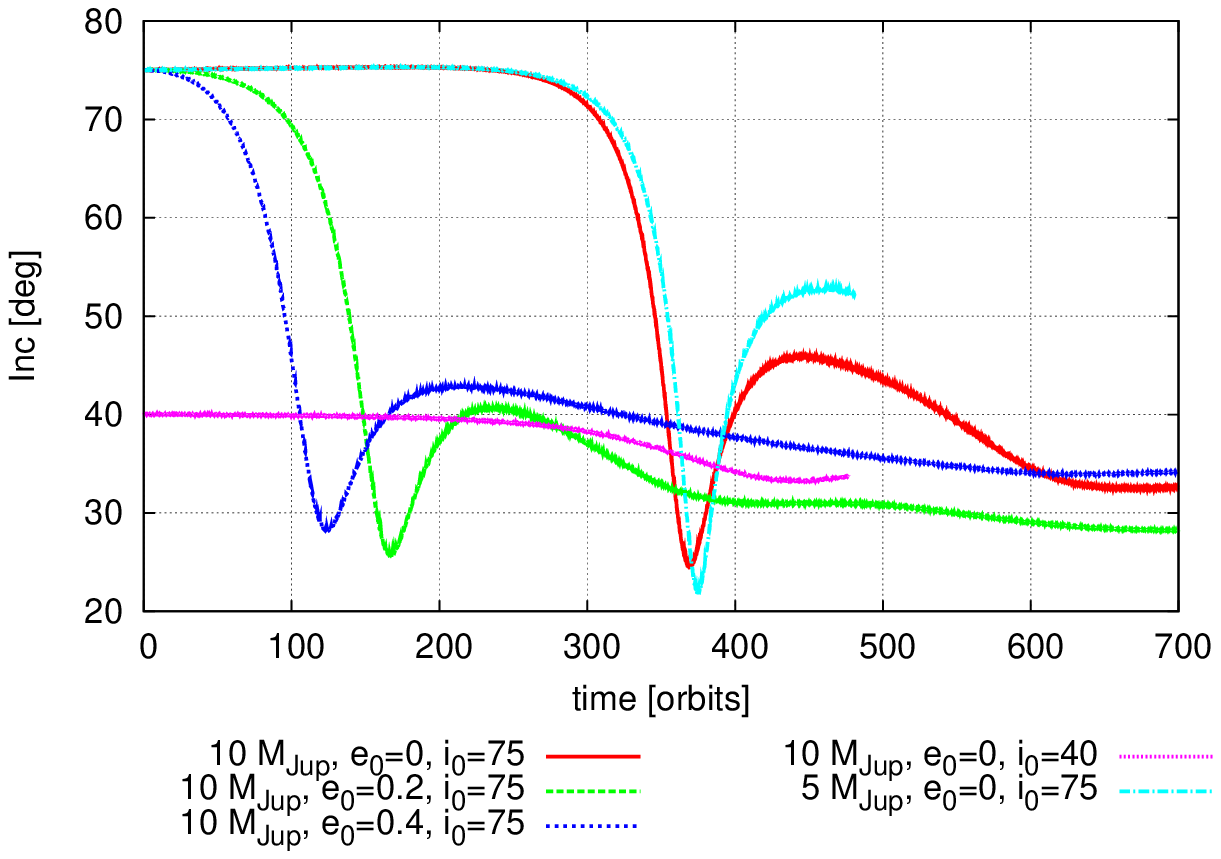}}
 \resizebox{\hsize}{!}{\includegraphics[width=0.9\linwx]{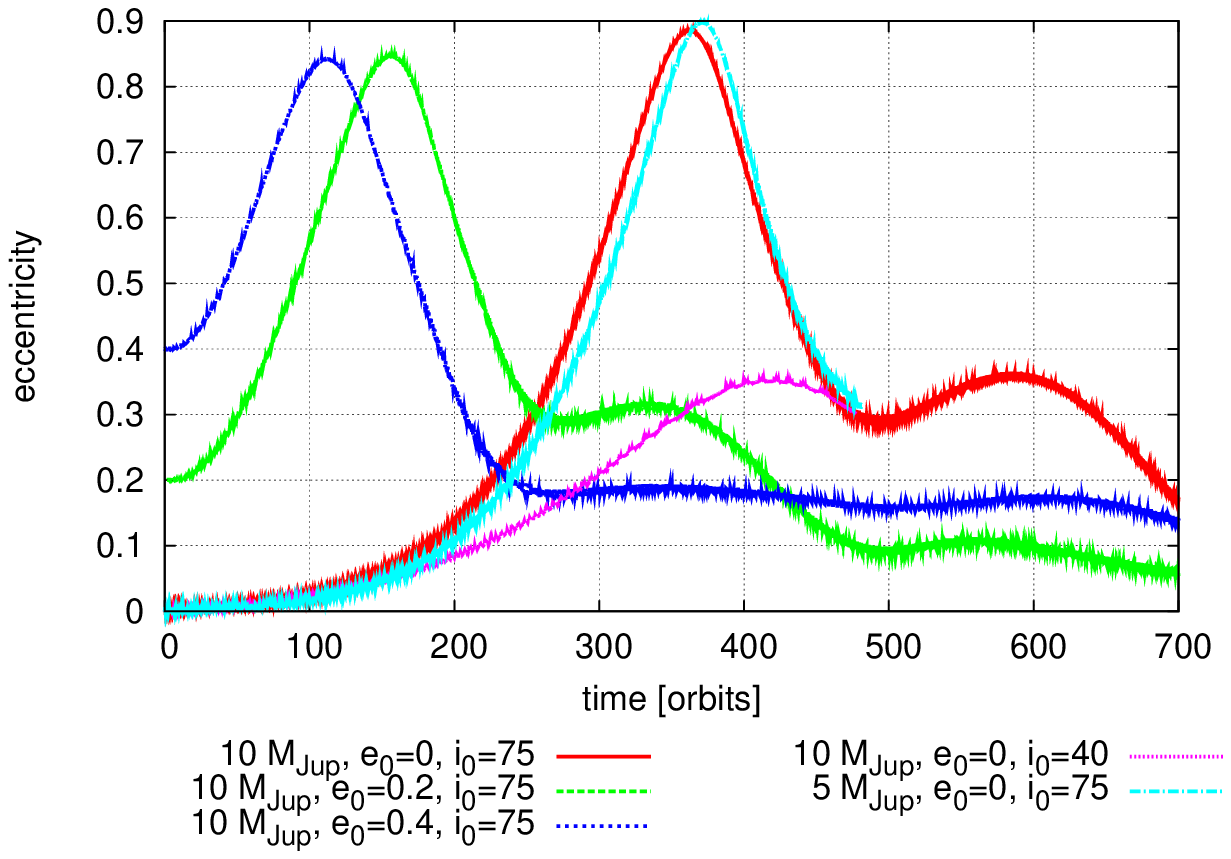}}
 \caption{Long-term evolution of planets with different inclinations, eccentricities, and masses in discs. The simulations are started from the equilibrium structures with fixed planets, where the planets are then released and allowed to move freely inside the disc. The top plot features the inclination of the planets, while the bottom plot shows the eccentricity of the planets. In the beginning the changes of eccentricity and inclination match the ones displayed in Figs.~\ref{fig:eccentricity} and \ref{fig:inclination}.
   \label{fig:kozai}
   }
\end{figure}

The long-term evolution of planets with different inclinations and eccentricities is displayed in Fig.~\ref{fig:kozai}. At the beginning of the evolution, the change of inclination and eccentricity matches those presented in Figs.~\ref{fig:eccentricity} and \ref{fig:inclination} for planets in fixed orbits. However, the evolution after the initial orbits is quite different from what was expected by the previous simulations.

In the $10M_{Jup}$, $e_0=0.0$, $i_0=75^\circ$ case, the inclination initially increases slightly with a rate that corresponds to the predicted rate (see also Fig.~\ref{fig:Inc40ev}). At the same time, the eccentricity of the planet increases and after about $250$ orbits it reaches $e\approx 0.25$. This eccentricity corresponds to inclination damping (Fig.~\ref{fig:inclination}), which is what happens in the evolution of the planet: the inclination drops. However, the eccentricity still increases at the same time, which was not predicted by the analysis of planets in fixed orbits (Fig.~\ref{fig:eccentricity}). The eccentricity then rises until a peak of $e\approx 0.9$, where it starts to drop again. At the same time, the inclination decrease stops and the inclination starts to rise again. As soon as the eccentricity has dropped to $e\approx 0.4$, the inclination starts to decrease again.

This exchange of inclination and eccentricity is representative of
the Kozai mechanism, introduced initially to describe the evolution of
a highly inclined asteroid perturbed by Jupiter \citep{1962P&SS....9..719L, 1962AJ.....67..591K}. A similar Kozai
mechanism affects the orbits of highly inclined planets with respect
to a disc \citep{2010MNRAS.404..409T, 2013MNRAS.428..658T}. For inclinations above a critical value, the gravitational force
exerted by the disc on the planet produces Kozai cycles where the
eccentricity of the planet can be pumped to high values, in antiphase
with its inclination. We note that the Kozai mechanism is visible
in the given computation time because of the high mass values
considered in our study ($5$ and $10 M_{Jup}$), comparable to the total mass
of the disc ($0.01 M_\odot$). Indeed high masses induce faster dynamical
evolution.

When the planet starts at a larger initial eccentricity ($e_0=0.2$ or
$e_0=0.4$), the general behaviour is similar as can be seen in
Fig.~\ref{fig:kozai}, but the first Kozai cycle occurs earlier. Circular orbits at high inclination constitute an unstable
equilibrium of the secular dynamics, so the evolution at zero initial
eccentricity remains for a while close to the separatrix associated with
the equilibrium \citep{2007Icar..191..469L}. The Kozai effect does not act for initial inclinations smaller
than a critical value ($i_0<~40^\circ$ in the restricted problem of
\citet{1962AJ.....67..591K}). We therefore also display a planet with $i_0=40^\circ$,
and show that the eccentricity and inclination oscillations are
significantly reduced.

Even if eccentricity can be pumped to high values, the Kozai mechanism
only postpones the alignment with the disc and the circularization of
the orbit induced by damping forces of the disc on the planet (given
in Figs.~\ref{fig:eccentricity} and \ref{fig:inclination}). As clearly shown by the evolution of the planet with
$e_0=0.2$, Kozai cycles repeat with reduced intensity. The drop of
inclination is much larger than the raise of inclination, after the
eccentricity increase/decrease cycle. These results are in agreement
with \citet{2013MNRAS.428..658T}, showing that low-mass planets would
remain on inclined and eccentric orbits over the disc lifetime, while
higher mass planets would align and circularize. We also illustrate in
Fig.~\ref{fig:kozai} the influence of the planet mass by considering a planet of $5
M_{Jup} (e=0.0)$: the eccentricity value reached during the second cycle
of inclination increase (at $450$ orbits) is higher for the $5 M_{Jup}$ planet, as expected.

The effect of Kozai oscillations between a disc and planet was also stated in \citet{2013MNRAS.431.1320X} however, \citet{2013MNRAS.431.1320X} were not able to resolve a full Kozai cycle, probably because their mass-ratio between planet and disc is smaller than in our case. This shows that for $i_0>40^\circ$, the measure of the forces on a planet on a fixed orbit is not relevant. In this case, damped Kozai oscillations will govern the long-term evolution of the orbital parameters. This phenomenon can be of crucial importance for the study of the fate of planets scattered on high-inclination orbits.

\subsection{Fitting for e and i}
\label{subsec:fit}

In Figs.~\ref{fig:inclination} and \ref{fig:eccentricity} we provided the change of $di/dt$ and $de/dt$ for different planetary masses. In these plots, lines indicate a fit for these data points. We now present the fitting formulae, which depend on the planet mass $M_P$, the eccentricity $e_P$, and inclination $i_P$. The inclination $i_P$ used in the presented formulae is given in degrees, as is the resulting $di/dt$. As discussed in the previous section, these formulae are only relevant for $i_0<40^\circ$ where no complex cycles are observed. Therefore, in fitting our parameters we have ignored the data points corresponding to high inclinations, in particular the ones showing inclination increase. This applies to the fitting of inclination and eccentricity.

As can be seen in the figures, the results of the numerical simulations are all but smooth. Therefore, one should not expect the fit to be very accurate with simple functions. However, our goal is to catch the big picture, and to provide an acceptable order of magnitude of the effect of the disc on the inclination and eccentricity. In log scale, the data appear relatively close to an increasing power law of $i_P$ for small $i_P$, and a decreasing power law of $i_P$ for large $i_P$. Therefore, we base our fits on the general form for the damping rates
\begin{equation}
\mathcal{F}(i_P)=-\frac{M_{disc}}{0.01\,M_\star}\left(ai_P^{-2b}+ci_P^{-2d}\right)^{-1/2} \ ,
\label{eq:general}
\end{equation}
where $b$ is positive and $d$ is negative. This way, for small $i_P$, $\mathcal{F}(i_P)\approx i_P^{b}\left(M_{disc}/0.01\,M_\star\sqrt{a}\right)$, and for large $i_P$, $\mathcal{F}(i_P)\approx i_P^{d}\left(M_{disc}/0.01\,M_\star\sqrt{c}\right)$. The coefficients $a$, $b$, $c$, and $d$ depend on the planet mass and eccentricity, and are fitted to the data as follows. The damping rate also has to be linear dependent on the disc mass $M_{disc}/M_\star$, as our simulations linearly scale with the gas density.

\subsubsection{Eccentricity}

We do not want ($de/dt$) to tend to zero when $i_P$ tends to zero. A pure increasing power law of $i_P$ is inappropriate here. The damping function will be given by
\begin{equation}
\mathcal{F}_e(i_P)=-\frac{M_{disc}}{0.01\,M_\star}\left(a(i_P+i_D)^{-2b}+ci_P^{-2d}\right)^{-1/2}\ ,
\label{eq:FF_e}
\end{equation} 
where $i_D$ is a small inclination so that for $i_P\approx 0$, $de/dt\approx -\frac{M_{disc}}{0.01\,M_\star}\frac{i_D^b}{\sqrt{a}}$. We are using $i_D=\tilde{M}_p/3$ degrees in Eq.~\ref{eq:FF_e}, where $\tilde{M}_p=1000\,M_p/M_\star$ is the planet mass in Jupiter masses. For small eccentricities, it is expected that $e_P/(de/dt)=\tau_e$ is constant. This makes the coefficient $a$ proportional to $e_P^{-2}$. We find that $de/dt$ is well fitted by the above general form using the coefficients
\begin{eqnarray}
a_e(M_P,e_P) &= &80\,e_P^{-2}\, \exp \left(-e_P^2 \tilde{M}_p / 0.26 \right) 15^{\tilde{M}_p} \, \left(20 + 11\tilde{M}_p-\tilde{M}_p^2 \right)
 \nonumber \\
  b_e(M_P) &= &0.3\tilde{M}_p \nonumber \\
  c_e(M_P) &= &450+2^{\tilde{M}_p} \nonumber \\
  d_e(M_P) &= &-1.4+\sqrt{\tilde{M}_p}/6 \ .
\label{eq:F_e}
\end{eqnarray}
The second degree polynomial function of $\tilde{M}_p$ in the expression of $a$ is just a refinement, its value being between $30$ and $50$ for $1<\tilde{M_p}<10$. We note, however, that it is negative for $\tilde{M_p}>12$ so this expression only applies for $\tilde{M_p}<11$, but this covers the range of giant planets. To describe the change of eccentricity we add a second function ${\cal G}_e$, which describes the eccentricity increase for high-mass planets. The damping and excitation of $e_P$ are two different mechanisms that add on the planet, and one of them finally dominates, setting the sign of $de/dt$. The fits in Fig.~\ref{fig:eccentricity} are the added functions. 

For ${\cal G}_e$ we use the result of \citet{2001A&A...366..263P} who calculated the eccentricity excitation for $i_0=0^\circ$ high mass-planets. Our calculation is presented in Appendix~\ref{ap:eccentricity} and gives
\begin{equation}
{\cal G}_e|_{i=0} = 12.65\, \frac{M_P M_{disc}}{M_{\star}^2}\, e_P\ .
\end{equation}
Then, we find that this excitation decreases with $i$ as a Gaussian function, finally making
\begin{eqnarray}
\label{eq:gaue}
 {\cal G}_e (i_P,M_P,e_P) = 12.65 \, \frac{M_P M_{disc}}{M_{\star}^2}\, e_P \, \exp\left(-\left(\frac{(i_P/1^\circ)}{\tilde{M}_p}\right)^2\right) \ .
\label{eq:G_e}
\end{eqnarray}

In principle planets with $M_P < 5 M_{Jup}$ and $e<0.3$ do not show any signs of eccentricity increase and the Gaussian function should not be added in that case. However, the function is designed to scale with the planetary mass, so that lower mass planets are not affected by it. The change of eccentricity is then given by the sum of ${\cal F}_e$ and ${\cal G}_e$.

\subsubsection{Inclination}

In the case of inclination damping data, we notice that the decreasing power law dominates actually before the intersection with the increasing power law\,; thus, we multiply the term $ai_P^{-2b}$ in our general formula by a Gaussian function of $i_P$ centred on $0^\circ$, so that this term is not affected for small $i_P$ but vanishes more quickly than is natural. It allows our fitting formula to catch the peak of damping in inclination observed around $5$ to $20$ degrees in Figure~\ref{fig:inclination}. For small $i_P$, $di/dt$ should be close to linear in $i_P$, so the coefficient $b$ should be close to $1$. The damping function for inclination ${\cal F}_i$ is then given, in degrees per orbit, by
\begin{eqnarray}
 a_i(M_P,e_P) &= &1.5 \cdot 10^4 (2-3e_P){\tilde{M}_p}^3\nonumber \\
 b_i(M_P,e_P) &= &1+\tilde{M}_p e_P^2 /10 \nonumber \\
 c_i(M_P,e_P) &= &1.2 \cdot 10^{6}/\big[ (2-3e_P) (5+e_P^2 (\tilde{M}_p+2)^3) \big] \nonumber \\
 d_i(e_P) &= &-3+2e_P \nonumber \\
 g_i(M_P,e_P) &= &\sqrt{3\tilde{M}_p / (e_P+0.001)}\times 1^\circ \nonumber \\
\mathcal{F}_i(M_P,e_P,i_P) &= & -\frac{M_{disc}}{0.01\,M_\star}\left[ a_i\, \left(\frac{i_P}{1^\circ}\right)^{-2b_i}\exp(-(i_P/g_i)^2/2) \right.\\
 &  & \hspace{3.5cm} + \left.c_i\,\left(\frac{i_P}{40^\circ}\right)^{-2d_i}\ \right]^{-1/2} \ . \nonumber
\label{eq:F_i}
\end{eqnarray}
We note that the expression for coefficient $c_i$ is clearly not valid for $e>2/3$.

\begin{figure}
 \centering
 \resizebox{\hsize}{!}{\includegraphics[width=0.9\linwx]{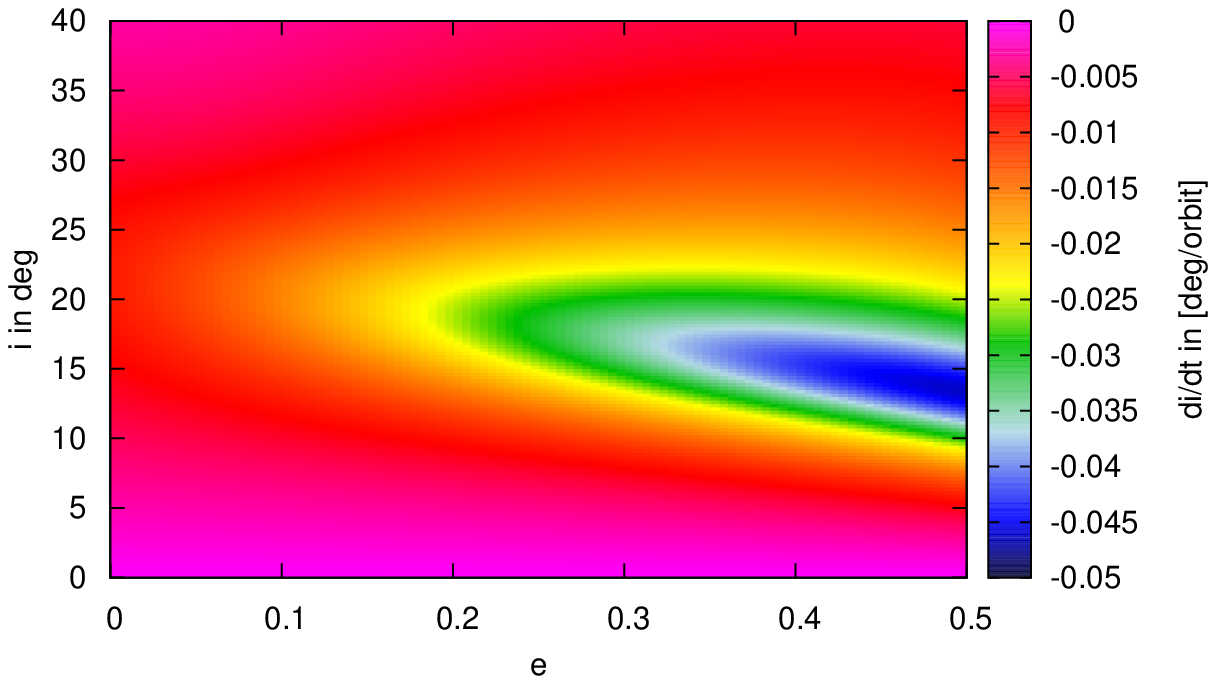}}
 \resizebox{\hsize}{!}{\includegraphics[width=0.9\linwx]{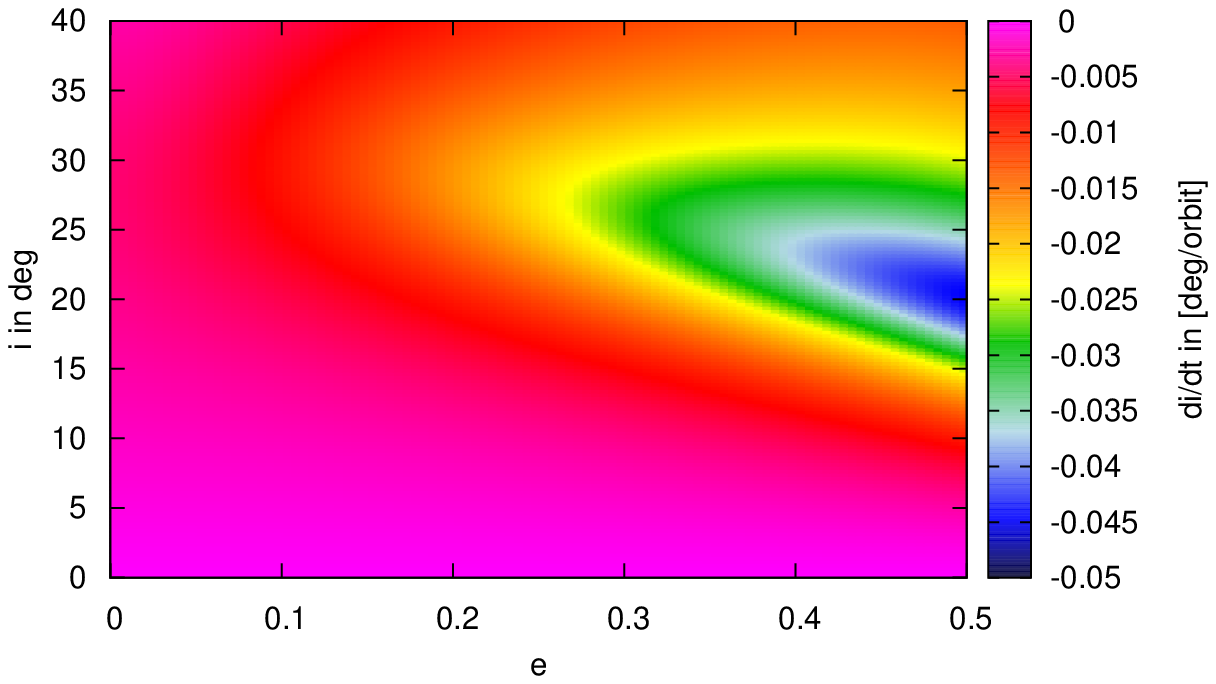}}
 \caption{{\bf top:} $di/dt$ for a $5M_{Jup}$ planet with different eccentricities and inclinations. The values of $di/dt$ have been determined with the formula given in Sect.~\ref{subsec:fit}. {\bf bottom:} same, but for $10M_{Jup}$. We note the different colour coding as the change is dependent on the planetary mass.
   \label{fig:2Ddidt}
   }
\end{figure}

\begin{figure}
 \centering
 \resizebox{\hsize}{!}{\includegraphics[width=0.9\linwx]{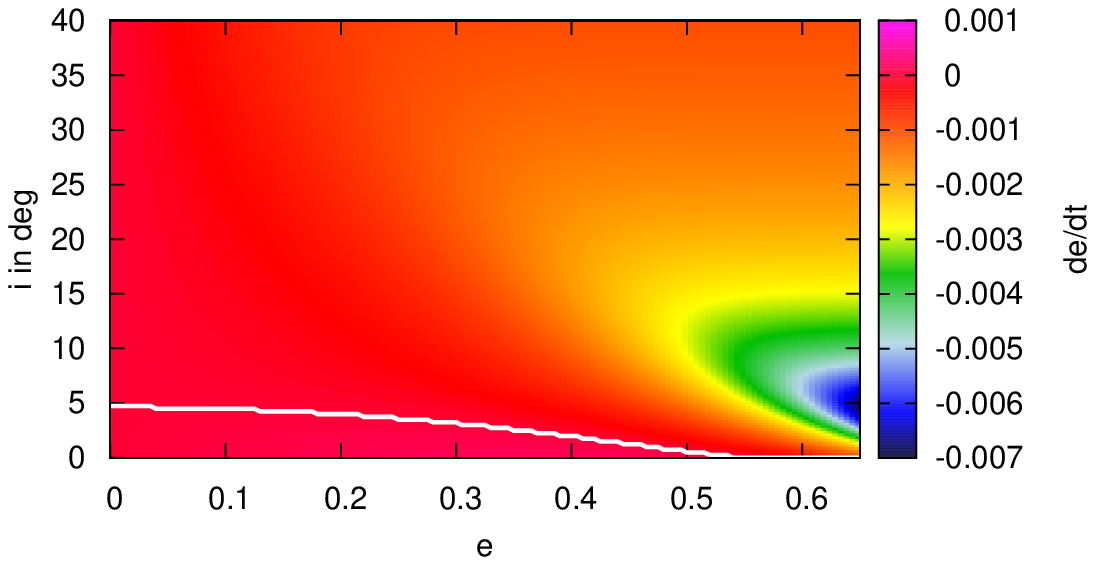}}
 \resizebox{\hsize}{!}{\includegraphics[width=0.9\linwx]{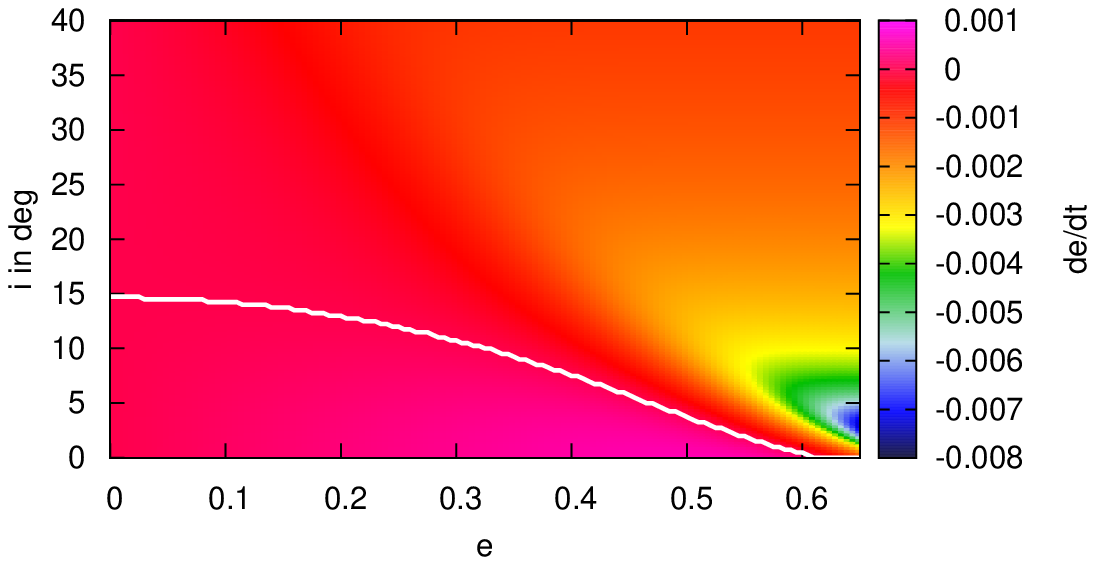}}
 \caption{{\bf top:} $de/dt$ for a $5M_{Jup}$ planet with different eccentricities and inclinations. The values of $de/dt$ have been determined with the formula given in Sect.~\ref{subsec:fit}. {\bf bottom:} same, but for $10M_{Jup}$. The white line in the figure indicates the transition between eccentricity increase and eccentricity damping. Below the white line, the eccentricity increases, above the line eccentricity decreases. We note the different colour coding as the change is dependent on the planetary mass.
   \label{fig:2Ddedt}
   }
\end{figure}

From our formulae for $de/dt$ and $di/dt$ we can now estimate how the eccentricity and inclination of a planet will evolve for all $e_P$ and $i_P$. In Fig.~\ref{fig:2Ddidt} the $di/dt$ for different inclinations and eccentricities for $5M_{Jup}$ and $10M_{Jup}$ according to the formulae is presented. In Fig.~\ref{fig:2Ddedt} the $de/dt$ for the same two planetary masses is plotted.

Figure~\ref{fig:2Ddidt} clearly indicates that the damping rate of inclination is highest for planets with a large eccentricity that are moderately inclined above the midplane ($i_P\approx 15^\circ$). The inclination damping rate indicates that planets that are scattered during the gas disc phase in orbits with moderate inclination ($i_P<40^\circ$), would lose their inclination well within the gas dispersal of the disc. 

As already indicated in Fig.~\ref{fig:eccentricity}, the eccentricity is always damped for high inclinations. For high planetary masses, the eccentricity of the planet can increase for low planetary inclinations because of interactions with disc. We find an eccentricity increase for both high-mass cases, but the increase of eccentricity declines with increasing eccentricity and inclination. Additionally, the threshold of $e_P$ and $i_P$ for which eccentricity can increase is larger for higher mass planets, which is indicated by the white line in Fig.~\ref{fig:2Ddedt} that represents the transition from eccentricity increase to decrease. Below the line eccentricity increases, above the line eccentricity decreases.

\section{Application to single-planet systems}
\label{sec:application}

\begin{figure}
 \centering
 \resizebox{\hsize}{!}{\includegraphics[width=0.9\linwx]{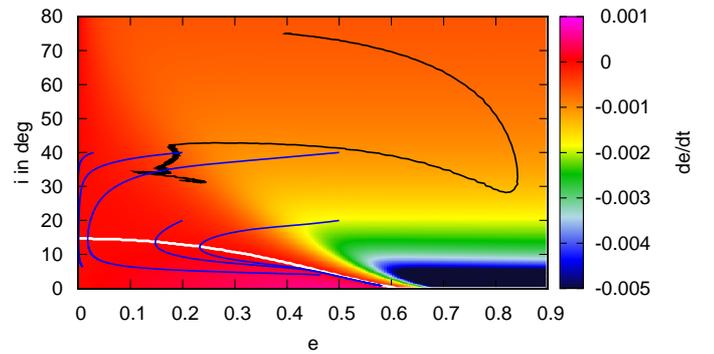}}
 \caption{Evolution of $e$ and $i$ of the $10 M_{Jup}$ planet (shown in Fig.~\ref{fig:kozai}) with $i_0=75^\circ$ and $e_0=0.4$ in the $e$-$i$ plane (black line). The background is the extended formula of the fit for $de/dt$ and the white line marks the transition between eccentricity increase and damping as in Fig.~\ref{fig:2Ddidt}. The blue lines indicate calculated trajectories of $10 M_{Jup}$ planets from the fitting formulae.
   \label{fig:Kozaimove}
   }
\end{figure}

The movement of a single planet in the disc can only be predicted if $i<40^\circ$ and $e<0.65$ as the planet would undergo a Kozai-oscillation for larger $i$. Additionally, the fitting formula might not be totally accurate for $e>0.5$, since our simulations only cover an eccentricity space of up to $e=0.4$. In Fig.~\ref{fig:Kozaimove} the trajectory of the $10 M_{Jup}$ planet with $i_0=75^\circ$ and $e_0=0.4$, which was shown in Fig.~\ref{fig:kozai} is displayed. This illustrates that the movement of the planet is a complex process as long as the Kozai-oscillations are still operational, but as soon as $i<40^\circ$, the planet loses inclination, which is then not converted back into eccentricity. The planet is damped towards midplane on a non-zero eccentricity. This non-zero eccentricity will actually hold in this case (see Section.~\ref{subsubsec:Eccentricity}).

A typical damping rate of $de/dt=0.001 /orbit$ would suggest that the planet will lose $\approx 0.085$ in eccentricity in the period of $10^4$ years. A typical damping rate of $di/dt = 0.01 deg/orbit$ indicates that the planet will lose $\approx 8.5^\circ$  of inclination in $10^4$ years.

The important parameters for the evolution of the orbit of a planet are the damping timescales $\tau_e=e/(de/dt)$ and $\tau_i=i/(di/dt)$. We find that $e/\mathcal{F}_e$ is much smaller than $i/\mathcal{F}_i$ for $i>10-20^\circ$, depending on the planet mass and eccentricity. Thus, planets scattered on highly inclined orbits will follow a certain pattern. While the inclination is damped slowly and still high, the eccentricity is damped to zero. After the inclination is damped further, the eccentricity of the planet can rise because of interactions with the disc (if $e$ is below the white line in Fig.~\ref{fig:Kozaimove}). Finally the inclination is damped to zero and the planet remains with a non-zero eccentricity. This is illustrated by the blue lines in Fig.~\ref{fig:Kozaimove} that correspond to calculated trajectories of $10M_{Jup}$ planets.

Nevertheless, this suggests that at the time of the disc dispersal, the favoured endstate for the planet's evolution is an eccentric orbit in midplane of the disc. This implies that the scattering process of inclined planets must have taken place after the gas is depleted or gone.

\section{Summary}
\label{sec:summary}

We have presented the evolution of eccentricity $e$ and inclination $i$ of high-mass planets ($M_P \geq 1M_{Jup}$) in isothermal protoplanetary discs. The planets have been kept on fixed orbits around the host star, and the forces from the disc acting onto the planet have been calculated. By using these forces, a change of $de/dt$ and $di/dt$ has been determined. 

Inclination and eccentricity are in general damped by the interactions with the disc. For $1M_{Jup}$ the damping rate of $e$ and $i$ is highest for only very small inclinations ($i_0\approx 3^\circ$), while the maximal damping rate is shifted to larger inclinations for more massive planets. As the more massive planets carve deeper gaps inside the disc, the damping interactions with the disc are reduced. But for larger inclinations, the planet can feel the full damping potential of the disc and is therefore damped in $e$ and $i$ at a faster rate.

There are two exceptions. The first is for low-inclination planets with a sufficient mass ($M_P > 4-5 M_{Jup}$). In this case, the interactions of the planet with the disc result in an increase of eccentricity of the planet, which has already been observed and studied \citep{2001A&A...366..263P, 2006A&A...447..369K}. However, our 3D results predict an increase of eccentricity for lower planetary masses than the previous studies.

The second exception arises for massive planets ($M_P \approx M_{disc}$, in our case for $M_P>5M_{Jup}$) on high initial inclinations ($i_0>40^\circ$). In the long-term evolution of the planet, eccentricity can increase, while inclination is damped and vice-versa. The planet undergoes a Kozai-cycle with the disc, but in time the oscillations of the planet in $e$ and $i$ diminish, as $e$ and $i$ get damped by the disc at the same time. The planet will end up in midplane through the interactions with the disc.

In Sect.~\ref{subsec:fit} we provided formulae for $di/dt$ and $de/dt$ for high-mass planets, which we fitted to the numerical hydrodynamical simulations. The formulae can now be used to calculate the long-term evolution of planetary systems during the gas phase of the disc with N-Body codes. However, we recommend not using the fitting formula, if the planetary eccentricity is $e>0.65$ and if $i>40^\circ$ (because of the Kozai interactions, a fit that can be used for the long-term evolution of planets is hard to predict).

In the end, the planet's inclination will be damped to zero. Low-mass planets ($M_P< 4-5 M_{Jup}$) will end up in circular orbits in the midplane of the disc, while higher mass planets ($M_P > 5M_{Jup}$) will pump their eccentricity to larger values because of interactions with the disc. This implies that the scattering process of inclined planets must have taken place after the gas is well depleted.

The influence of the gaseous protoplanetary disc on the inclination is also of crucial importance, if multiple planets are present in the disc that excite each other's inclination during their migration \citep{2009MNRAS.400.1373L, 2011MNRAS.412.2353L}. The influence of the disc on the long-term evolution of multi-body systems will be studied in a future paper.

\begin{acknowledgements}

B. Bitsch has been sponsored through the Helmholtz Alliance {\it Planetary Evolution and Life}. The work of A.-S. Libert is supported by an FNRS Postdoctoral Research Fellowship. The calculations were performed on systems of the Computer centre (ZDV) of the University of T\"ubingen and systems  operated by the ZDV on behalf of bwGRiD, the grid of the Baden W\"urttemberg state. We thank the referee Willy Kley for his useful and helpful remarks that improve the paper.

\end{acknowledgements}

\appendix

\section{Additional information on numerics}
\label{ap:numerics}

\begin{figure}
 \centering
 \resizebox{\hsize}{!}{\includegraphics[width=0.9\linwx]{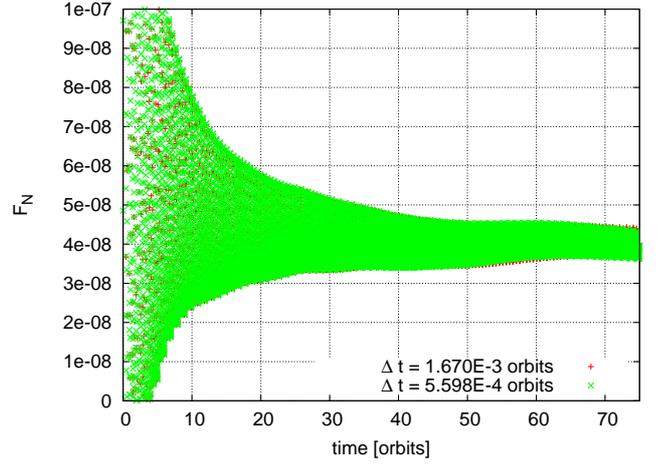}}
 \caption{$F_N$ for $5M_{Jup}$ planets with $i=75^\circ$ on circular orbits. The planet's feature different time steps, as indicated in the plot. $F_N$ has been averaged over $1$ running orbit. 
   \label{fig:FNtime}
   }
\end{figure}

In principle, a fast vertical movement (more than $1$ grid cell per timestep) through the grid could cause problems with the Fargo algorithm, as Fargo shifts the grid cells for several cells azimuthally and the effects of the planet on the gas might get corrupted. In Fig.~\ref{fig:FNtime} we present the evolution of the normal component of the disturbing force $F_N$ (which has been averaged over $1$ orbit) of planets with $i=75^\circ$ on circular orbits. The two simulations shown feature different time-step lengths. For the simulation with larger timestep, the planet moves through about one vertical grid cell in each time step. For the shorter timestep, three timesteps are needed to cover the vertical extent of one grid cell. The evolution of $F_N$ seems to be identical, indicating that the length of the timestep is not of crucial importance here also because we use a rotating frame so that the planet is on a fixed position inside the numerical grid where the Fargo algorithm does not shift grid cells for $r \approx a_P$. 

The smoothing of the planetary potential is crucial for avoiding singularities at the planet's location. In Sect.~\ref{subsec:pot} the numerical potential for the planets was introduced. Of crucial importance here is the smoothing length $r_{sm}$. A smaller smoothing length $r_{sm}$ leads to a deeper planetary potential. This in turn leads to a larger accumulation of mass at the planet's location, but this increase in density near the planet can be very high for large planets, especially in the isothermal case. This increase of density near the planet is unphysical, as normally the temperature and pressure gradients should stop the accumulation of gas at some point, which is not possible in the isothermal case. In this situation, the density can become so large that the gradients of density near the planet become too steep and the timestep inside the code collapses down to very small values, which makes an integration over several orbits impossible. We therefore return to the $\epsilon$-potential for the $10M_{Jup}$ planet.

\begin{figure}
 \centering
 \resizebox{\hsize}{!}{\includegraphics[width=0.9\linwx]{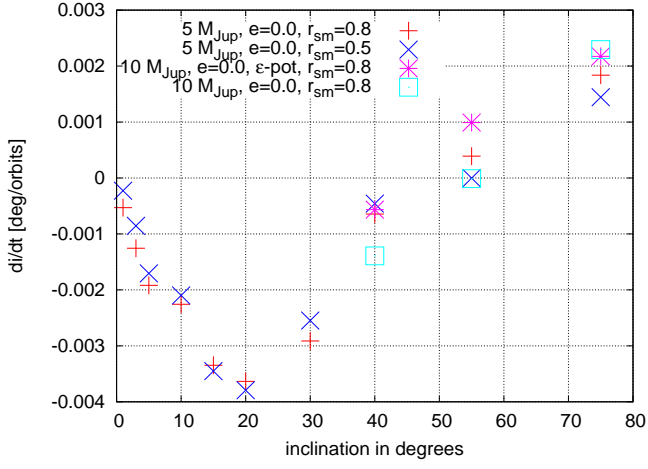}}
 \caption{Change of inclination $di/dt$ for $5M_{Jup}$ and $10M_{Jup}$ planets in circular orbits for two different smoothing length for the planetary potential. The cubic potential is used for all displayed simulations, unless stated otherwise.
   \label{fig:5MJuppot}
   }
\end{figure}

In Fig.~\ref{fig:5MJuppot} we present the inclination damping for $5M_{Jup}$ planets in circular orbits for two different smoothing lengths, $r_{sm}=0.8$ and $r_{sm}=0.5$. Changing the planetary potential seems to influence the damping of inclination by up to $\pm 15\%$, but the general trend is the same. Even with a deeper planetary potential, the inclination of a planet seems to increase for large initial inclinations. The main difference seems to be that no inclination increase can be observed for the $i_0=55^\circ$ case with a smoothing length of $r_{sm}=0.5$. This has also been observed for the $10M_{Jup}$ planet where the difference between the depth of the two potentials is supposed to be stronger (as we change from the $\epsilon$ to the cubic potential), but the trend is the same as for the $5M_{Jup}$ planet. For $i_0=75^\circ$ the inclination seems to increase for models of planets in fixed orbits for both $5M_{Jup}$ and $10M_{Jup}$. We therefore conclude that the general trend is conserved regardless of the chosen planetary potential and smoothing length.

\begin{figure}
 \centering
 \resizebox{\hsize}{!}{\includegraphics[width=0.9\linwx]{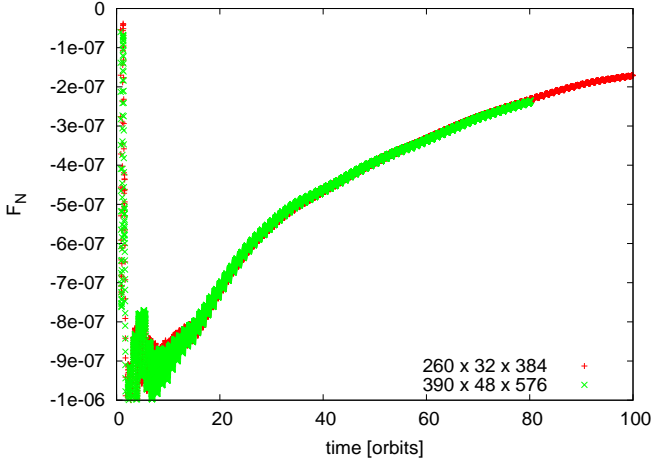}}
 \caption{$F_N$ for a $10M_{Jup}$ planet with $i=3^\circ$ in a $e_0=0.4$ orbit. $F_N$ has been averaged over $1$ running orbit. 
   \label{fig:FNres}
   }
\end{figure}

In order to find the sufficient numerical resolution for our simulations of inclination damping, we have performed several resolution tests. In Fig.~\ref{fig:FNres} we present the results of these tests. The plotted quantity $F_N$ has been averaged over $1$ running orbit. Keep in mind that $F_N$ has been averaged over $40$ orbits to determine the change of inclination in the end. The simulations feature a $10M_{Jup}$ planet with $i=3^\circ$, so it is well embedded inside the disc. The numerical resolution of the grid has been changed from $260 \times 32 \times 384$ to $390 \times 48 \times 576$. As the crucial force $F_N$ for inclination damping gives the same results for both resolutions, we use the lower resolution for our simulation with confidence.

\section{Eccentricity and inclination of the disc}
\label{ap:eidisc}

To determine the eccentricity and inclination of the disc, we take a mass-weighted average of the eccentricity of all grid cells. To compute the eccentricity we take the specific total energy (in mass units)
\begin{equation}
 E_{tot, spec.} = - \frac{G M_\star}{r} + \frac{1}{2} \mathbf{v}^2 \ ,
\end{equation}
where $\mathbf{v}$ is the velocity vector of a given grid cell and $r=\sqrt{x^2+y^2+z^2}$ the radial component towards the grid cell. The total energy is given by 
\begin{equation}
 E_{tot} = - \frac{G M_\star}{2a} \ ,
\end{equation}
where $a$ is the semi-major axis towards the grid cell. With that, we can compute $a$
\begin{eqnarray}
 - \frac{G M_\star}{r} + \frac{1}{2} \mathbf{v}^2 &=& - \frac{G M_\star}{2a} \nonumber \\ 
 \Rightarrow \quad a &=& -\frac{G M_\star}{2} / \left( \frac{1}{2} \mathbf{v}^2 - \frac{G M_\star}{r} \right) \ .
\end{eqnarray}
With $a$ we can now compute the eccentricity $e$ of each grid cell:
\begin{equation}
 L_{spec.} = \sqrt{G M_\star a (1-e^2)} \quad \Rightarrow \quad e = \sqrt{1-\frac{L_{spec.}^2}{G M_\star a}} \ ,
\end{equation}
where $L_{spec.} = \mathbf{r} \times \mathbf{v}$ is the specific angular momentum of each grid cell. To get an estimate of the eccentricity of the disc, we make a mass-weighted average of the eccentricity of each grid cell (averaged in azimuthal and polar coordinates) in order to get $e_{disc} (r)$:
\begin{equation}
 e_{disc} (r) = \frac{\Sigma m_{\theta \phi} e_{\theta \phi}}{\Sigma m_{\theta \phi}} \ ,
\end{equation}
where $m_{\theta \phi}$ is the mass of the grid cell.

Because we use spherical coordinates $r$, $\theta$, $\phi$ for the inclination, we have to transform $\mathbf{L}_{spec.}$ first into Cartesian coordinates in order to calculate the mass average. This has to be done because each product $\mathbf{L}=\mathbf{r} \times  \mathbf{v}$ is given in a different local coordinate system of each grid cell, but for the average the angular momentum vectors should always be in the same coordinate frame. The angular momentum vector is given in the two coordinate systems by 
\begin{eqnarray}
 \mathbf{L} &=& \mathbf{L}_r \mathbf{u}_r +  \mathbf{L}_\theta \mathbf{u}_\theta + \mathbf{L}_\phi \mathbf{u}_\phi \nonumber \\
 \mathbf{L} &=& \mathbf{L}_x \mathbf{u}_x +  \mathbf{L}_y \mathbf{u}_y + \mathbf{L}_z \mathbf{u}_z \ ,
\end{eqnarray}
where
\begin{eqnarray}
 \mathbf{u}_r &=& \sin \theta \cos \phi \mathbf{u}_x + \sin \theta \sin \phi \mathbf{u}_y + \cos \theta \mathbf{u}_z \nonumber \\
 \mathbf{u}_\theta &=& \cos \theta \cos \phi \mathbf{u}_x + \cos \theta \sin \phi \mathbf{u}_y - \sin \theta \mathbf{u}_z \nonumber \\
 \mathbf{u}_\phi &=& - \sin \phi \mathbf{u}_x + \cos \theta \mathbf{u}_y \ ,
\end{eqnarray}
with the angles $\theta$ and $\phi$ of the grid cell, which differ for each grid cell. This gives us for $\mathbf{L}$ in Cartesian coordinates
\begin{eqnarray}
 \mathbf{L} &=& (\cos \theta \cos \phi \mathbf{L}_\theta  - \sin \phi \mathbf{L}_\phi ) \mathbf{u}_x \nonumber \\
 &+& ( \cos \theta \sin \phi \mathbf{L}_\theta + \cos \phi \mathbf{L}_\phi ) \mathbf{u}_y \nonumber \\
 &-& \sin \theta \mathbf{L}_\theta \mathbf{u}_z \ .
\end{eqnarray}
For the inclination of the disc we now take a mass-averaged specific angular momentum (averaged in polar and azimuthal direction)
\begin{equation}
 \mathbf{L}_{av.} (r) = \frac{\Sigma \mathbf{L}_{spec., c} m_c}{\Sigma m_c} \ ,
\end{equation}
where the subscript $c$ denotes the grid cell number, and $m_c$ the corresponding mass of the grid cell. Now we can compute the angle between $\mathbf{L}_{av.}$ and the $z$-axis, which gives us the averaged inclination at each ring of the disc.

\section{Increase of eccentricity}
\label{ap:eccentricity}

We follow \citet{2001A&A...366..263P} to calculate the maximum value of eccentricity increase for high-mass planets, as it is given by $A_g$ in Eq.~\ref{eq:gaue}. In \citet{2001A&A...366..263P} the increase of eccentricity is calculated through the growth rates of the modes of the Lindblad resonance, which is given by
\begin{equation}
 \gamma = \frac{1}{4 {\cal J}} \frac{d{\cal J}}{dt} \ ,
\end{equation}
where
\begin{equation}
 {\cal J} = - \frac{1}{2} M_P e_P^2 \sqrt{(GM_\star)r_P} - \frac{1}{2} \int \Sigma e_d^2 r^3 \Omega \  dr d\phi \ ,
\end{equation}
where $e_d$ is the disc's eccentricity and $r_P$ the planetary distance to star. The integral basically gives the disc mass, which is comparable to the planet's mass, but as $e_d$ is much smaller than $e_P$ (see Fig.~\ref{fig:eidisc10MJup}), the term concerning the disc eccentricity is much smaller than the term concerning the planetary eccentricity. We therefore choose to neglect it in our estimate of the eccentricity increase. We then get
\begin{equation}
 \frac{d{\cal J}}{dt} = - M_P \dot{e}_P e_P \sqrt{(GM_\star)r_P} \ ,
\end{equation}
which leads to 
\begin{equation}
 \gamma = \frac{- M_P \dot{e}_P e_P \sqrt{(GM_\star)r_P} }{-4(\frac{1}{2} M_P e_P^2 \sqrt{(GM_\star)r_P})} = \frac{1 \dot{e}_P}{2e_P} \ .
\end{equation}
As also
\begin{eqnarray}
 \frac{\gamma}{\omega} &= &\frac{M_{disc} M_P}{M_\star^2} \left( \frac{r_P}{r} \right)^8 \nonumber \\
 &\times &\frac{9\pi \left[ (r\bar{e_d} - \frac{2r}{3} \frac{d\bar{e_d}}{dr}) \frac{r}{r_P} - \frac{21}{4} e_P (1+ \frac{5}{7} (r_P/r)^2) \right]^2}{e_P^2 + \int 2 \pi \Sigma e_d^2 r^3 \Omega dr / (M_P \omega r_P^2)} \ ,
\end{eqnarray}
where we set $e_d=0.0$ and $\bar{e_d}=0.0$ because we are only interested in a first order estimate of the eccentricity increase from the disc. With $\omega = \sqrt{GM_\star / r_P^3}$ we find for $\dot{e}_P=2e_P \gamma$
\begin{equation}
 \dot{e}_P = (12.65 M_P M_{disc}/M_{\star}^2) e_P = \mathcal{G}_e|_{i=0} \ ,
\end{equation}
which gives the increase of eccentricity for a planet orbiting in the midplane of the disc, which fits quite well with the results of our simulations (Fig.~\ref{fig:eccentricity}).

\bibliographystyle{aa}
\bibliography{Inc}
\end{document}